\documentclass[10pt]{article}

\usepackage{jheppub}
\usepackage{esint,shuffle,psfrag}
 \usepackage[utf8]{inputenc}
\usepackage{hyperref}

\usepackage{graphicx} % inclusion des figures
\usepackage{float}
\usepackage{color}
\usepackage[toc,page]{appendix} 
\usepackage{enumerate}
\usepackage{latexsym,amsthm,amssymb,amsfonts,amsbsy,amsmath}
\usepackage{amsmath}
\usepackage{dsfont}
\usepackage{amsfonts}
\usepackage{amssymb}
\usepackage{mathrsfs}
\usepackage{bbm}
\usepackage{mathtools}

\usepackage{amsthm}
\usepackage{bm} % symboles mathématiques gras
\usepackage{bbm} % font mathbb gras
\usepackage{slashed} % notation de Feynman
\usepackage{cancel}
\usepackage{xcolor}
\usepackage{tikz}
\usetikzlibrary{arrows,matrix,snakes}
\usetikzlibrary{decorations.markings,shapes.geometric,shapes.misc} 
\usepackage{tikz-cd}

\usepackage{pdflscape}
\usepackage{scalefnt}

\definecolor{lightred}{RGB}{255,127,127}
\definecolor{lightgreen}{RGB}{127,255,127}
\definecolor{lightblue}{RGB}{127,127,255}
\definecolor{linkcolor}{rgb}{0,0,0.6}
\numberwithin{equation}{section}

\usepackage{tabularx} % gestion avancée des tableaux

\usepackage{fancyhdr} % entêtes et pieds de pages personnalisés

\theoremstyle{plain}

\def\res{\mathop{\text{res}\,}}

\newcommand{\Gh}{\widehat{\Gamma}}

\newcommand{\vp}{\varphi}

\newcommand{\p}{\partial}
\newcommand{\g}{\mathfrak{g}}
\newcommand{\ti}[1]{_{\bm{\underline{#1}}}}

\newcommand{\dd}{\text{d}}

\newcommand{\C}{\mathbb{C}}

\newcommand{\R}{\mathbb{R}}
\newcommand{\D}{\mathbb{D}}

\newcommand{\Pc}{\mathcal{P}}
\newcommand{\Id}{\text{Id}}

\newcommand{\s}{\sigma}

\newcommand{\Rc}{\mathcal{R}}
\newcommand{\Lc}{\mathcal{L}}

\newcommand{\Hc}{\mathcal{H}}
\newcommand{\Mc}{\mathcal{M}}

\newcommand{\ze}{\zeta}
\newcommand{\Q}{\mathcal{Q}}

\newcommand{\Ww}[1]{I_{\text{W}\hspace{-1pt}\text{Z}}\bigl[#1\bigr]}

\newcommand{\Cc}{\mathcal{C}}

\newcommand{\Oc}{\mathcal{O}}

\newcommand{\diag}{\text{diag}}
\newcommand{\T}[2]{T^{#1}_{{\color{white}#1}#2}}

\DeclareFontEncoding{LS1}{}{}
\DeclareFontSubstitution{LS1}{stix}{m}{n}
\DeclareSymbolFont{stixsymbols}{LS1}{stixscr}{m}{n}
\SetSymbolFont{stixsymbols}{bold}{LS1}{stixscr}{b}{n}
\DeclareMathSymbol{\kay}{\mathalpha}{stixsymbols}{"6B}

\def\res{\mathop{\text{res}\,}}

\definecolor{myGreen}{rgb}{0.0,0.4,0.0}

\title{\Large New Integrable Coset Sigma Models}

\author[a]{Gleb Arutyunov,}
\author[a]{Cristian Bassi}
\author[a]{and Sylvain Lacroix}

\affiliation[a]{II. Institut f\"ur Theoretische Physik, Universit\"at Hamburg, Luruper Chaussee 149, 22761
Hamburg, Germany\\
Zentrum  f\"{u}r  Mathematische  Physik,  Universit\"{a}t  Hamburg,  Bundesstrasse  55,  20146  Hamburg,Germany}

\emailAdd{gleb.arutyunov@desy.de}  
\emailAdd{cristian.bassi@desy.de} 
\emailAdd{sylvain.lacroix@desy.de}

\abstract{By using the general framework of affine Gaudin models, we construct a new class of integrable sigma models. 
They are defined on a coset of the direct product of $N$ copies of a Lie group over some diagonal subgroup and they depend on $3N-2$ free parameters.
For $N=1$ the corresponding model coincides with the well-known symmetric space sigma model.
Starting from the Hamiltonian formulation,  we derive the Lagrangian for the $N=2$ case and show that it 
admits a remarkably simple form in terms of the classical $\Rc$-matrix underlying the integrability of these models. 
We conjecture that a similar form of the Lagrangian holds for arbitrary $N$.
Specifying our general construction 
to the case of $SU(2)$ and $N=2$, and eliminating one of the parameters,   
we find a new three-parametric integrable model with the manifold $T^{1,1}$ as its target space.  
We further comment on the connection of our results with those existing in the literature.  

}

\begin{document}
\begin{flushright}\small{ZMP-HH/20-19}\end{flushright}

\maketitle

\flushbottom

\section{Introduction}
The remarkable recent progress in applying integrability techniques to the models of the gauge-string correspondence has given further impetus to the study 
of possible origins and general properties of integrable field theories.  In particular, it was shown in  \cite{Vicedo:2017cge}  that many classical 
integrable field theories can be viewed as specific realisations of dihedral affine Gaudin models, associated with an untwisted affine Kac-Moody algebra
supplied with an action of a dihedral group. A characteristic feature of these field theories is  that their Poisson structure is non-ultralocal and the known examples 
include, for instance, the principal chiral model and its integrable deformations, affine Toda field theories, etc. In fact, one can turn the logic around and ask whether it is possible to employ the general framework  of  dihedral affine Gaudin models that have built-in integrability   to construct novel examples of integrable field theories. This is precisely what has been exploited in the recent work 
\cite{Delduc:2018hty,Delduc:2019bcl}, where this approach was used to construct a new class of integrable sigma models  
that couple together an arbitrary number of principal chiral model fields on the same Lie group.

The aim of the present work is to make a further step towards the exploration of the panorama of  affine Gaudin models.
Namely, we will show how to construct integrable sigma models on a coset of the direct product of $N$ copies of an arbitrary real semi-simple Lie group $G$ over some diagonal subgroup,
generalising the standard symmetric space construction corresponding to the $N=1$ case (the existence of such models has been conjectured in \cite{Delduc:2018hty,Delduc:2019bcl}).

To explain the logic of our construction, we recall that affine Gaudin models are naturally defined in the Hamiltonian setting. 
The Poisson structure on the phase space given in terms  of Takiff currents admits different realisations: the one we are interested in here is in terms of canonical fields parametrising $N$ copies of the cotangent bundle $T^*G$   \cite{Delduc:2019bcl}. Following \cite{Vicedo:2017cge},
the Hamiltonian of the model is chosen to ensure that the dynamics takes the form of the zero curvature condition for a Lax connection.
Further, we define  $G^{(0)}$ as a subgroup of $G$ invariant under the action  of an involutive automorphism 
$\sigma$ and embed it into $G^N$ as the diagonal subgroup $G^{(0)}_{\text{diag}}\subset G^N$. The current group of $G^{(0)}_{\text{diag}}$ 
acts on the phase space by gauge transformations and, in particular, on $G^N$-valued fields by multiplications from the right.  This action is Hamiltonian and it gives rise to a moment map which, under a certain condition, is a first-class constraint.  As a next step, we perform the standard Hamiltonian reduction by fixing the value of the moment map to zero.  The corresponding locus of the phase space 
should be then factorised by the action  of the local $G^{(0)}_{\text{diag}}$ leaving us with a model on the reduced phase space corresponding to a coset $G^N/G^{(0)}_{\text{diag}}$. This is how the coset construction  is performed in the Hamiltonian setting. Note that the affine Gaudin model, {\it i.e.} its Poisson structure and its Hamiltonian, 
 depends on  $3N-2$ free parameters which are all encoded  in its so-called twist function.  
 
 Since we are primarily interested in the Lagrangian description of the coset model, we need to perform the inverse Legendre transform and this constitutes the most non-trivial technical part. In particular,  to integrate out the momenta, we first derive their Lagrangian description in terms of time derivatives of group elements and this derivation involves solving the Hamiltonian constraint in an explicit manner.  For simplicity we restrict ourselves to the case $N=2$ and 
 obtain a sigma model action with a Wess-Zumino term that couples two group elements $g_1, g_2\in  G$, see (\ref{Eq:Action}). This action 
 exhibits a gauge $G^{(0)}_{\text{diag}}$-symmetry acting on $g_1$ and $g_2$ by right multiplication, therefore rendering the model to be defined on the coset $G\times G/G^{(0)}_{\text{diag}}$.  
 The emergence of gauge symmetry is natural, 
 as  in the process of eliminating the momenta we have only solved the Hamiltonian constraint, postponing factorisation by the local  $G^{(0)}_{\text{diag}}$.
 As a result,  $G^{(0)}_{\text{diag}}$ shows up at the Lagrangian level as a gauge symmetry. Finally, we also present the Lagrangian form of the Lax connection 
 that guarantees integrability of the sigma model equations of motion.

The sigma model part of the action we obtain is given by the sum of quadratic combinations of currents with coefficients depending on $3N-2=4$ free parameters.
What is remarkable is that this action can be recast in a very simple form involving the classical 
$\Rc$-matrix that underlies the integrability of the model. Moreover, this form of the action admits a straightforward generalisation (\ref{Eq:ActionRef}) to any $N$,
which we verify for $N=3$.
 
Having obtained these general results, it is interesting to consider some limits or to focus on some particular models. First, it appears that for the $N=2$ case 
one can define a scaling limit in which one of the four parameters decouples leaving behind a three-parameter $(\lambda,\lambda_1,\lambda_2)$ 
family  of integrable models. 
We then observe that at the particular point  $\lambda_1=\lambda_2=\lambda$ the corresponding action coincides with the one of the 
Guadagnini-Martellini-Mintchev model \cite{Guadagnini:1987ty} on the homogeneous space $G\times G/G^{(0)}_{\text{diag}}$. This model defines 
a two-dimensional conformal field theory and its integrability has already been established in~\cite{Bardakci:1996gs}. We then show that the general Lax connection specified to this model
acquires a very simple form.

Finally, in the above $N=2$ three-parameter model we specify $G=SU(2)$ and $G^{(0)}=U(1)$ and 
obtain a gauged sigma model on the coset $SU(2)\times SU(2)/U(1)$. Fixing the gauge by putting one of the Cartan angles to zero, we obtain the gauge-fixed
action from which we read off the sigma-model metric and the $B$-field. The metric turns out to coincide with the three-parameter 
family of metrics on the $T^{1,1}$ manifolds
\begin{equation}
\nonumber
ds^2 =  \lambda_1^2(\text{d}\theta_1^2 + \sin^2\theta_1 \, \text{d}\phi^2_1) + \lambda_2^2(\text{d}\theta_2^2 + \sin^2\theta_2 \, \text{d}\phi^2_2) + \lambda^2(\text{d}\psi + \cos\theta_1 \, \text{d}\phi_1 +  \cos\theta_2 \, \text{d}\phi_2)^2\, .
\end{equation}
Particularly interesting configurations of parameters correspond to $\lambda_1^2=\lambda_2^2=3\lambda^2/2$ 
and $\lambda_1=\lambda_2=\lambda$. In the first case, we get a sigma model on an Einstein manifold, in the second case we obtain the already mentioned 
conformal model which, in particular, was used in \cite{PandoZayas:2000he} to construct pure NS-NS supergravity solutions.

Although integrability of the geodesic flow on $T^{1,1}$ has been already established in~\cite{Benvenuti:2005ja,deCellis:2012ai}, what follows from our consideration is that 
the sigma model on a generic three-parameter $T^{1,1}$ is integrable and we present the corresponding Lax connection. For integrability to hold the presence of the $B$-field
\begin{equation}
\nonumber
B= \lambda^2 (\dd\psi + \cos\theta_1\,\dd\phi_1) \wedge (\dd\psi + \cos\theta_2\,\dd\phi_2)
\end{equation}
is crucial. In particular, changing  the overall coefficient $\lambda^2$ to any  other value destroys integrability. 
To support this claim, we  consider an isometry-preserving setting where 
the $B$-field is allowed with an arbitrary coefficient. In order to probe (non-)integrable properties of this generalised model,  we 
reduce the sigma-model equations to a mechanical system by plugging in them the so-called spinning string ansatz, in the spirit of     
\cite{Kim:2003vn,Wang:2005baa,Basu:2011di,Basu:2011fw,Rigatos:2020hlq} where spinning (or wrapped) strings on $T^{1,1}$ were studied.  At the end we obtain a coupled system of differential 
equations for the two angle coordinates $\theta_1$ and $\theta_2$. We then observe that only when the coefficient of the $B$-field is $\lambda^2$, 
the equations for $\theta_1$ and $\theta_2$ decouple (separate) and can be integrated by quadrature.  In any other case there is no decoupling and most probably 
the corresponding dynamical system exhibits a non-integrable behaviour, similar to what has been found in \cite{Basu:2011di,Basu:2011fw,Rigatos:2020hlq}.
 
The paper is organised as follows. In the next section we construct the coset models in the Hamiltonian formulation. In section 3 we derive the action of the coset sigma model for $N=2$, rewrite this action in a new form involving the classical $\Rc$-matrix and discuss further generalisations for arbitrary $N$. 
We also consider a limiting case where one of the parameters is scaled away and at a special point in the parameter space we find 
the match of the corresponding model with the conformal model of
Guadagnini, Martellini and Mintchev. Section 4 is devoted to integrable sigma models on $T^{1,1}$ manifolds. 
We relegate some technical details to three appendices.  

\section{Construction of the models in the Hamiltonian formulation}  \label{Sec:HamiltonianForm}

Let $G$ be a connected semi-simple real Lie group, $\sigma$ an involutive automorphism of $G$ and $G^{(0)}$ the subgroup of fixed-points of $\sigma$. Our goal in this section is to construct integrable $\sigma$-models on $G^N/G^{(0)}_{\text{diag}}$, where $N$ is a positive integer and $G^{(0)}_{\text{diag}}=\lbrace (h,\cdots,h), \, h\in G^{(0)} \rbrace$. As we shall see, they will be more precisely obtained as models on $G^N$ with a $G^{(0)}_{\text{diag}}$ gauge symmetry. For $N=1$, the construction will yield the standard $\sigma$-model on the symmetric space $G/G^{(0)}$, which is well known to be integrable.

The formalism we will use to construct these integrable field theories is the one of dihedral affine Gaudin models, introduced in~\cite{Vicedo:2017cge}, which is naturally defined in the Hamiltonian formulation of classical field theories. In this context, the phase space of the models will consist of canonical fields on the cotangent bundle $T^*G^N$, together with a first-class constraint encoding the $G^{(0)}_{\text{diag}}$ gauge symmetry. The approach followed in this section is reminiscent of the one developed in~\cite{Delduc:2018hty,Delduc:2019bcl} to construct integrable $\sigma$-models on $G^N$, without taking quotients by a subgroup. A reformulation of these models on the quotient $G^{N+1}/G_{\text{diag}}$, closer to the approach used here, was proposed in~\cite{Lacroix:2019xeh}.

We will start by reviewing the phase space of canonical fields on one copy of $T^*G$ in subsection \ref{SubSec:TStarG}. We will then proceed to define the structure of the models as dihedral affine Gaudin models in subsection \ref{SubSec:AGM}. In subsection \ref{Sec:Hamiltonian}, we will define the Hamiltonian of these field theories as well as the constraint corresponding to their $G^{(0)}_{\text{diag}}$ gauge symmetry. Subsection \ref{Sec:Momentum} will be concerned with space-time symmetries of the models and in particular with the determination of a simple condition ensuring their relativistic invariance. In subsection \ref{SubSec:Integrability} we will prove that these models are integrable. Finally, in subsection \ref{Sec:Panorama} we describe the panorama of models obtained through this construction and in particular discuss their defining parameters.

\subsection{Phase space of canonical fields on \texorpdfstring{$\bm{T^*G}$}{T*G}}
\label{SubSec:TStarG}

\paragraph{Conventions and notation.} \label{Sec:Conventions}  Let $\mathfrak{g}$ be the Lie algebra of the group $G$. We denote by $\kappa$ the opposite of the Killing form of $\mathfrak{g}$: it defines a non-degenerate ad-invariant bilinear form on $\g$, which is definite positive if $G$ is compact. Let us also fix a basis of $\mathfrak{g}$, which we will denote by $(I_a)_{a \in \{1,\ldots,n\}}$. We will indicate the dual of this basis with respect to $\kappa$ by $(I^a)_{a \in \{1,\ldots,n\}}$. In the following we will often make use of the so called split quadratic Casimir of $\mathfrak{g}$, which is defined as the following element:
\begin{equation} \label{Eq:Casimir}
C_{\underline{\mathbf{12}}} = I_a \otimes I^a
\end{equation}
in $\mathfrak{g} \otimes \mathfrak{g}$ and which is independent of the choice of basis (here and in the following, we use the standard tensorial notations $\underline{\mathbf{i}}$). From the definition of $C_{\underline{\mathbf{12}}}$ and the ad-invariance of the bilinear form $\kappa$, one checks that
\begin{equation*}
\kappa\ti{2} \bigl( C\ti{12}, X\ti{2} \bigr) = X \qquad \text{ and } \qquad \bigl[ C\ti{12}, X\ti{1} + X\ti{2} \bigr] = 0
\end{equation*}
for all $X\in\g$.\\

Let $\sigma$ be an involutive automorphism of $G$ and $G^{(0)} \subset G$ be the subgroup of fixed-points of $\sigma$. It induces an involutive automorphism of the Lie algebra $\mathfrak{g}$, which we also call $\sigma$ by a slight abuse of notation. As $\sigma$ is of order two, it has eigenvalues $+1$ and $-1$. We define the corresponding eigenspaces
\begin{equation*}
\mathfrak{g}^{(0)} = \{X \in \mathfrak{g}: \sigma(X) = X\}, \,\,\,\,\,\,\,\,\,\, \text{and} \,\,\,\,\,\,\,\,\,\, \mathfrak{g}^{(1)} = \{X \in \mathfrak{g}: \sigma(X) = -X\}\,.
\end{equation*}
These eigenspaces form a $\mathbb{Z}_2$-gradation of $\mathfrak{g}$: $\mathfrak{g} = \mathfrak{g}^{(0)} \oplus \mathfrak{g}^{(1)}$, with
\begin{equation*}
[\mathfrak{g}^{(0)},\mathfrak{g}^{(0)}] \subset \mathfrak{g}^{(0)}, \,\,\,\,\,\,\,\,\,\, [\mathfrak{g}^{(0)},\mathfrak{g}^{(1)}]  \subset\mathfrak{g}^{(1)} \,\,\,\,\,\,\,\,\,\, \text{and} \,\,\,\,\,\,\,\,\,\, [\mathfrak{g}^{(1)},\mathfrak{g}^{(1)}] \subset \mathfrak{g}^{(0)}\,.
\end{equation*}
The converse is also true, \textit{i.e.} given a $\mathbb{Z}_2$-gradation of $\mathfrak{g}$, there is a unique automorphism $\sigma$ which leaves $\mathfrak{g}^{(0)}$ invariant and acts on $\mathfrak{g}^{(1)}$ as multiplication by $-1$. In particular, $\g^{(0)}$ is a subalgebra of $\g$, which is the Lie subalgebra corresponding to the subgroup $G^{(0)}$ in $G$.

In the following we will use the notation $X^{(i)}$ to indicate the component of an element $X\in\g$ in $\mathfrak{g}^{(i)}$, $i \in \{0,1\}$. More precisely, if we call $\pi^{(0)}= (\text{Id}+\sigma)/2$ and $\pi^{(1)}= (\text{Id}-\sigma)/2$ the projectors on $\mathfrak{g}^{(0)}$ and $\mathfrak{g}^{(1)}$ respectively, we then have $X^{(i)} = \pi^{(i)}X$, for $X = X^{(0)} + X^{(1)}$ a generic element of $\mathfrak{g}$.

It is a standard result that the automorphism $\sigma$ preserves the bilinear form $\kappa$. Hence, $\mathfrak{g}^{(0)}$ and $\mathfrak{g}^{(1)}$ are orthogonal with respect to the bilinear form $\kappa$, or, in other words, $\kappa\left(\mathfrak{g}^{(0)},\mathfrak{g}^{(1)}\right) = 0$. Moreover, the split quadratic Casimir \eqref{Eq:Casimir} satisfies
\begin{equation*}
\sigma_{\underline{\mathbf{1}}}\sigma_{\underline{\mathbf{2}}}C_{\underline{\mathbf{12}}} = C_{\underline{\mathbf{12}}}\,.
\end{equation*}
For $i\in\lbrace 0,1 \rbrace$, we define the projection $C_{\underline{\mathbf{12}}}^{(ii)}= \pi_{\underline{\mathbf{1}}}^{(i)}\pi_{\underline{\mathbf{2}}}^{(i)}C_{\underline{\mathbf{12}}}$ of the split quadratic Casimir on $\g^{(i)}\otimes\g^{(i)}$. Let us note that the orthogonality of $\g^{(0)}$ and $\g^{(1)}$ implies that $\pi_{\underline{\mathbf{1}}}^{(i)}\pi_{\underline{\mathbf{2}}}^{(j)}C_{\underline{\mathbf{12}}} = \delta_{ij} C_{\underline{\mathbf{12}}}^{(ii)}$, for $i,j\in\lbrace 0,1 \rbrace$. Moreover, we have
\begin{equation}\label{Eq:CasGrad}
\kappa\ti{2} \bigl( C^{(ii)}\ti{12}, X\ti{2} \bigr) = X^{(i)}, \qquad \forall\, X\in\g\,.
\end{equation}

\paragraph{Canonical fields on $\bm{T^*G}$.} \label{Sec:Canonical} Let us consider canonical fields depending on a single space coordinate $x \in \mathbb{D}$ and taking values in the cotangent bundle $T^*G$. In this paper we fix $\mathbb{D}$ to be either the real line $\mathbb{R}$ or the circle $S^1$ and choose boundary conditions for the fields accordingly. 

Since $T^*G\simeq  G \times \mathfrak{g}$,
these fields can be described by a pair of fields $(g,X): \mathbb{D} \to G \times \mathfrak{g}$, which encode the coordinate and momentum fields respectively. Being a cotangent bundle, $T^*G$ has a natural Poisson bracket. Therefore, fields with values in $T^*G$ form the phase space of an Hamiltonian field theory. In terms of the fields $g$ and $X$, the Poisson bracket is given by:
\begin{subequations} \label{Eq:PoissongX}
\begin{align}
\{g_{\underline{\mathbf{1}}}(x),g_{\underline{\mathbf{2}}}(y)\} &= 0\,, \\
\{X_{\underline{\mathbf{1}}}(x),g_{\underline{\mathbf{2}}}(y)\} &= g_{\underline{\mathbf{2}}}(x) C_{\underline{\mathbf{12}}}\delta_{xy}\,, \\
\{X_{\underline{\mathbf{1}}}(x),X_{\underline{\mathbf{2}}}(y)\} &=  [C_{\underline{\mathbf{12}}},X_{\underline{\mathbf{1}}}(x)]\delta_{xy}\,,
\end{align}
\end{subequations}
where $C_{\underline{\mathbf{12}}}$ is the split quadratic Casimir \eqref{Eq:Casimir} of $\mathfrak{g}$ and $\delta_{xy} = \delta(x-y)$ is the Dirac delta-distribution.

\paragraph{Current $\bm{j(x)}$ and momentum.} Let us define the following $\mathfrak{g}$-valued current:
\begin{equation*}
j(x) = g^{-1}(x)\partial_xg(x)\,.
\end{equation*}
From \eqref{Eq:PoissongX}, it satisfies the Poisson brackets
\begin{subequations}  \label{Eq:Poissonj}
\begin{align}
\{g_{\underline{\mathbf{1}}}(x),j_{\underline{\mathbf{2}}}(y)\} &= 0\,, \\
\{j_{\underline{\mathbf{1}}}(x),j_{\underline{\mathbf{2}}}(y)\} &= 0\,, \\
\{X_{\underline{\mathbf{1}}}(x),j_{\underline{\mathbf{2}}}(y)\} &=  [C_{\underline{\mathbf{12}}},j_{\underline{\mathbf{1}}}(x)]\delta_{xy} -C_{\underline{\mathbf{12}}}\delta'_{xy}\,,
\end{align}
\end{subequations}
where $\delta'_{xy} = \p_x \delta(x-y)$ is the derivative of the Dirac delta-distribution with respect to $x$.

Let us also consider the quantity
\begin{equation} \label{Eq:Momentum}
\mathcal{P}_{G} = \int_{\mathbb{D}} \text{d}x \ \kappa(j(x),X(x))\,.
\end{equation}
From \eqref{Eq:PoissongX} and \eqref{Eq:Poissonj}, one can check that its Hamiltonian flow generates the spatial derivatives on both $g(x)$ and $X(x)$:
\begin{equation*}
\{\mathcal{P}_{G},g(x)\} = \partial_x g(x) \hspace{30pt} \text{and} \hspace{30pt} \{\mathcal{P}_{G},X(x)\} = \partial_x X(x)\,.
\end{equation*}
Hence, it is the momentum of the phase space.

\paragraph{Wess-Zumino term and current $\bm{W(x)}$.} \label{Sec:WZ} For this paragraph, let us consider the field $g$ to also depend explicitly on a time coordinate $t \in \mathbb{R}$ (in the Hamiltonian formulation, this time dependence is implicitly defined by the choice of a Hamiltonian). Let us further extend the space-time $\mathbb{D}\times\mathbb{R}$ (with coordinates $(x,t)$) to a 3-dimensional manifold $\mathbb{B}$ with boundary $\p\mathbb{B}=\mathbb{D}\times\mathbb{R}$ (parametrised by coordinates $(x,t,\xi)$) and let us consider an extension of the field $g$ to $\mathbb{B}$ (which restricts to the initial field $g$ on $\p\mathbb{B}$). The Wess-Zumino term of $g$ is then defined as~\cite{Wess:1971yu,Novikov:1982ei,Witten:1983ar}
\begin{equation}
I_{\text{WZ}}[g] = \iiint_{\mathbb{B}} \dd x \,\dd t \, \dd \xi \; \kappa\Bigl( \bigl[ g^{-1} \p_x g, g^{-1} \p_t g \bigr], g^{-1} \p_\xi g \Bigr)\,.
\end{equation}
Up to the addition of a constant term, it does not depend on the choice of extension of $g$ from $\mathbb{D}\times\mathbb{R}$ to $\mathbb{B}$. It is a standard result that the 3-form $\kappa\Bigl( \bigl[ g^{-1} \p_x g, g^{-1} \p_t g \bigr], g^{-1} \p_\xi g \Bigr) \, \dd x \wedge \dd t \wedge \dd \xi$ is closed and thus locally exact. Therefore, the Wess-Zumino term can be rewritten, at least locally, as a 2-dimensional integral on $\p\mathbb{B}=\mathbb{D}\times\mathbb{R}$, which takes the form 
\begin{equation} \label{Wesszumino}
I_{\text{WZ}}[g] = \iint_{\mathbb{D} \times \mathbb{R}} \text{d}x\,\text{d}t \ \kappa(W,g^{-1}\partial_tg)\,,
\end{equation}
where $W$ is a $\g$-valued current depending on the coordinate fields in $g$ and their spatial derivatives. We will not need here the precise definition of $W$ and refer for instance to \cite{Delduc:2019bcl} for more details.\\

In the Hamiltonian formalism, this current can be seen as a $\g$-valued local observable $W(x)$ on the phase space of canonical fields on $T^*G$. One can then show that it satisfies the following Poisson bracket with the fields $g$, $X$ and $j$ introduced above:
\begin{subequations}\label{Eq:PoissonW}
\begin{equation}\label{Eq:PoissonW1}
\{g_{\underline{\mathbf{1}}}(x),W_{\underline{\mathbf{2}}}(y)\} = 0\,, \hspace{40pt} \{j_{\underline{\mathbf{1}}}(x),W_{\underline{\mathbf{2}}}(y)\} = 0
\end{equation}
and
\begin{equation} \label{Eq:PoissonW2}
\{X_{\underline{\mathbf{1}}}(x),W_{\underline{\mathbf{2}}}(y)\} + \{W_{\underline{\mathbf{1}}}(x),X_{\underline{\mathbf{2}}}(y)\} =  [C_{\underline{\mathbf{12}}},W_{\underline{\mathbf{1}}}(x) - j_{\underline{\mathbf{1}}}(x)]\delta_{xy}\,.
\end{equation}
\end{subequations}
Moreover, let us note that it satisfies the following orthogonality property:
\begin{equation} \label{Eq:Orthogonal}
\kappa\bigl(j(x),W(x)\bigr) = 0\,.
\end{equation}

\subsection{Definition of the models as realisations of affine Gaudin models}
\label{SubSec:AGM}

In this section, we define the models that we will consider in this article as realisations of dihedral affine Gaudin models (AGM), following~\cite{Vicedo:2017cge}. We will not review here the complete construction of dihedral AGM and will instead restrict ourselves to the definition and description of the main ingredients that are useful to construct these models: their sites, their twist function and their Gaudin Lax matrix.

The adjective dihedral used above refers to certain equivariance properties under an action of the dihedral group $D_{2T}$ ($T\in\mathbb{Z}_{\geq 1}$) which are satisfied by the twist function and the Gaudin Lax matrix of the models~\cite{Vicedo:2017cge}. These properties have to do with reality conditions and with the choice of a $\mathbb{Z}_T$-grading of the Lie algebra $\g$. For the models that we are considering in this article, we have $T=2$. The corresponding choice of $\mathbb{Z}_2$-grading $\g^{(0)}\oplus\g^{(1)}$ of $\g$ is then given by the choice of an involutive automorphism $\sigma$, as described in the previous section. We will come back to the equivariance properties encoding the dihedrality at the end of this subsection.

\paragraph{Sites, levels and twist function.} Following the formalism and terminology of \cite{Vicedo:2017cge}, let us consider a dihedral AGM with $N \in \mathbb{Z}_{\geq 1}$ real sites of multiplicity two, whose positions will be denoted by $z_r$ with $r \in \{1,\ldots,N\}$ and will be supposed to be non zero ($z_r \in \mathbb{R}^*$). Since each site $z_r$ is of multiplicity two, it is associated with two constant numbers $\ell_{r,0} \in \mathbb{R}$ and $\ell_{r,1} \in \mathbb{R}^*$, called the levels. Altogether this data specifies the twist function $\varphi(z)$ of the model, which depends on an auxiliary complex parameter $z \in \mathbb{C}$, called the spectral parameter. This function takes the following form \cite{Vicedo:2017cge}:
\begin{equation}\label{Eq:Twist}
\varphi(z) = \frac{1}{2}\sum_{r=1}^N\sum_{p=0}^1\sum_{k=0}^1\frac{(-1)^{k}\ell_{r,p}}{((-1)^{k}z-z_r)^{p+1}}\,.
\end{equation}
The sum over $p\in\lbrace 0,1 \rbrace$ in this expression and thus the presence of double poles at $z_r$ reflect the fact that the sites of the model are of multiplicity two. Moreover, the sum over $k\in \lbrace 0,1 \rbrace$ and the factors $(-1)^k$ encode the $T=2$ dihedrality of the model.

In the rest of this article, we will suppose that the levels $\ell_{r,0}$ satisfy the following additional hypothesis, which for reasons to be explained later we call the first-class condition:
\begin{equation}\label{Eq:SumLevels}
\sum_{r=1}^N \ell_{r,0} = 0\,.
\end{equation}
As we shall see in subsection \ref{Sec:First-class}, this condition will be necessary to ensure that the models that we construct possess a gauge symmetry.

\paragraph{Takiff currents and phase space.} To each site $z_r$, $r\in\lbrace 1,\cdots,N \rbrace$, of the model is attached two $\g$-valued fields $\mathcal{J}_{r,[0]}(x)$ and $\mathcal{J}_{r,[1]}(x)$, called Takiff currents . These are observables on the phase space of the model, which satisfy the following Poisson bracket, determined by the choice of levels $\ell_{r,p}$:
\begin{subequations} \label{Eq:PoissonJ}
\begin{align}
\{\mathcal{J}_{r,[0] \underline{\mathbf{1}}}(x),\mathcal{J}_{s,[0] \underline{\mathbf{2}}}(y)\} &= \delta_{rs}\left([C_{\underline{\mathbf{12}}},\mathcal{J}_{r,[0] \underline{\mathbf{1}}}(x)]\delta_{xy} - \ell_{r,0} C_{\underline{\mathbf{12}}} \delta'_{xy}\right), \\
\{\mathcal{J}_{r,[0] \underline{\mathbf{1}}}(x),\mathcal{J}_{s,[1] \underline{\mathbf{2}}}(y)\} &= \delta_{rs}\left([C_{\underline{\mathbf{12}}},\mathcal{J}_{r,[1] \underline{\mathbf{1}}}(x)]\delta_{xy} - \ell_{r,1} C_{\underline{\mathbf{12}}} \delta'_{xy}\right), \\
\{\mathcal{J}_{r,[1] \underline{\mathbf{1}}}(x),\mathcal{J}_{s,[1] \underline{\mathbf{2}}}(y)\} &= 0\,.
\end{align}
\end{subequations}
So far, we did not specify what is the phase space of the model: this requires discussing the distinction between a formal AGM and its realisations. The phase space of the formal AGM underlying the present construction simply consists of configurations of the Takiff currents $\mathcal{J}_{r,[p]}(x)$ ($r\in\lbrace 1,\cdots,N \rbrace$ and $p\in\lbrace 0,1 \rbrace$), equipped with the Poisson bracket \eqref{Eq:PoissonJ}. Taking a realisation of this AGM consists of considering a more general phase space, describing configurations of fields $\phi_i(x)$ with a certain Poisson bracket, such that there exist well-chosen combinations $\mathcal{J}_{r,[p]}(x)$ of the fields $\phi_i(x)$ that satisfy the Takiff brackets \eqref{Eq:PoissonJ}. The construction of the formal AGM can then be completly transfered to this new phase space, thus yielding an integable field theory with observables on this space.

In the present case, we will consider a particular realisation of this AGM, whose phase space consists of canonical fields on the cotangent bundle $T^*G^N$. We described the phase space of canonical fields on one copy of $T^*G$ in subsection \ref{SubSec:TStarG}: we will use the notations and conventions introduced in this subsection to describe the fields on $T^*G^N$. In particular, these fields can be encoded into $N$ $G$-valued fields $g_1(x),\cdots,g_N(x)$ and $N$ $\g$-valued fields $X_1(x),\cdots,X_N(x)$, which are the equivalents of the fields $g(x)$ and $X(x)$ introduced in subsection \ref{SubSec:TStarG} for one copy of $T^*G$ and which then satisfy $N$ independent copies of the Poisson bracket \eqref{Eq:PoissongX}. Similarly, we introduce currents $j_r(x)$ and $W_r(x)$, $r\in\lbrace 1,\cdots,N \rbrace$, as the equivalent of the currents $j(x)$ and $W(x)$ of subsection \ref{SubSec:TStarG}. Let us then define
\begin{subequations}\label{Eq:TakiffJ}
\begin{align}
\mathcal{J}_{r,[0]}(x) &= X_r(x) + \frac{\ell_{r,0}}{2} j_r(x) + \frac{\ell_{r,0}}{2} W_r(x)\,, \label{Eq:KM} \\
\mathcal{J}_{r,[1]}(x) &= \ell_{r,1} \, j_r(x)\,.
\end{align}
\end{subequations}
It is a standard result (see for instance~\cite{Delduc:2019bcl}) that these satisfy the Takiff brackets \eqref{Eq:PoissonJ}, as can be checked directly from the brackets \eqref{Eq:PoissongX}, \eqref{Eq:Poissonj} and \eqref{Eq:PoissonW}. Thus, one can construct a realisation of the AGM considered above in the phase space of canonical fields on $T^*G^N$.

\paragraph{Gaudin Lax matrix.} Let $\mathfrak{g}^{\mathbb{C}}$ denote the complexification of $\mathfrak{g}$. The other fundamental piece needed for the construction of the model is the so-called Gaudin Lax matrix. It is defined as the following $\mathfrak{g}^{\mathbb{C}}$-valued field \cite{Vicedo:2017cge}:
\begin{equation}\label{Eq:Gaudin}
\Gamma(z,x) = \frac{1}{2}\sum_{r=1}^N\sum_{p=0}^1\sum_{k=0}^1\frac{(-1)^{k}\sigma^k\mathcal{J}_{r,[p]}(x)}{((-1)^{k}z-z_r)^{p+1}}\,.
\end{equation}
In this expression, the $T=2$ dihedrality of the model is encoded in the sum over $k\in\lbrace 0,1 \rbrace$ and the presence of the involutive automorphism $\sigma$. This is how the choice of $\sigma$ and thus the choice of the subgroup $G^{(0)}$ enters the definition of the model as AGM. From \eqref{Eq:PoissonJ}, one can compute the Poisson bracket of the Gaudin Lax matrix:
\begin{align}\label{Eq:PoissonG}
\{\Gamma_{\underline{\mathbf{1}}}(z,x),\Gamma_{\underline{\mathbf{2}}}(w,y)\} &= [\mathcal{R}^0_{\underline{\mathbf{12}}}(z,w),\Gamma_{\underline{\mathbf{1}}}(z,x)]\delta_{xy} - [\mathcal{R}^0_{\underline{\mathbf{21}}}(w,z),\Gamma_{\underline{\mathbf{2}}}(w,x)]\delta_{xy}  \nonumber \\
&\hspace{130pt} - \left(\mathcal{R}^0_{\underline{\mathbf{12}}}(z,w)\varphi(z) + \mathcal{R}^0_{\underline{\mathbf{21}}}(w,z)\varphi(w)\right)\delta'_{xy}\,,
\end{align}
with $\mathcal{R}^0_{\underline{\mathbf{12}}}$ given by
\begin{equation} \label{Eq:DefRMatrix}
\mathcal{R}^0_{\underline{\mathbf{12}}}(z,w) = \frac{1}{2}\sum_{k=0}^1 \frac{\sigma_{\underline{\mathbf{1}}}^k C_{\underline{\mathbf{12}}}}{w - (-1)^{k}z}\,,
\end{equation}
where we recognise the standard $\mathcal{R}$-matrix twisted by the automorphism $\sigma$. In particular, it satisfies the classical Yang-Baxter equation:
\begin{equation} \label{Eq:CYBE}
[\mathcal{R}^0_{\underline{\mathbf{12}}}(z_1,z_2),\mathcal{R}^0_{\underline{\mathbf{13}}}(z_1,z_3)] + [\mathcal{R}^0_{\underline{\mathbf{12}}}(z_1,z_2),\mathcal{R}^0_{\underline{\mathbf{23}}}(z_2,z_3)] + [\mathcal{R}^0_{\underline{\mathbf{32}}}(z_3,z_2),\mathcal{R}^0_{\underline{\mathbf{13}}}(z_1,z_3)] = 0\,.
\end{equation}

\paragraph{Dihedrality.} As mentioned at the beginning of this subsection, the AGM that we are considering here possesses certain equivariance properties under the dihedral group $D_4$. Let us now discuss these properties.

The general dihedral group $D_{2T}$ contains the cyclic group $\mathbb{Z}_{T}$ as a subgroup. Recall that for the models considered in this article, we have $T=2$: the corresponding cyclic group $\mathbb{Z}_2$ acts on the complex plane by multiplication by $-1$ and on the Lie algebra $\g$ by the involutive automorphism $\sigma$, which we extend to the complexification $\g^\C$ by $\C$-linearity. One checks from their expressions \eqref{Eq:Twist} and \eqref{Eq:Gaudin} that the twist function and the Gaudin Lax matrix are equivariant 1-forms with respect to these actions, \textit{i.e.} that
\begin{equation} \label{Eq:Equivariance}
\sigma(\Gamma(z,x)) = -\,\Gamma(-z,x) \hspace{30pt} \text{ and } \hspace{30pt} \varphi(z) = -\varphi(-z)\,.
\end{equation}
Let us note that the sums over $k\in\lbrace 0,1 \rbrace$ and the presence of the factors $(-1)^k$ and $\sigma^k$ in equations \eqref{Eq:Twist} and \eqref{Eq:Gaudin} are crucial for the above conditions to hold.\\

In addition to the cyclic group $\mathbb{Z}_T$, the dihedral group $D_{2T}$ contains an order two cyclic group $\mathbb{Z}_2$ (which is not to be confused with the $\mathbb{Z}_2$ group discussed above, which arises since we have $T=2$ in the case considered in this article). The equivariance properties corresponding to this $\mathbb{Z}_2$ subgroup encode the reality conditions of the model. It acts on the complex plane by conjugation $z\mapsto\bar z$ and on the complexified Lie algebra $\g^\C$ by the antilinear involutive automorphism $\tau$, defined such that the real form $\g$ is the subalgebra of fixed points of $\tau$. One checks that the automorphisms $\sigma$ and $\tau$ of $\g^\C$ satisfy the dihedrality confition $\sigma\circ\tau = \tau\circ\sigma$: the group generated by $\sigma$ and $\tau$ is thus isomorphic to the direct product $\mathbb{Z}_2 \times \mathbb{Z}_2$, which is the dihedral group\footnote{For a general $T$ (\textit{i.e.} when we have $\sigma$ of order $T$), the dihedrality condition reads $\sigma\circ\tau=\tau\circ\sigma^{-1}$ and the dihedral group $D_{2T}$ has the structure of a semi-direct product $\mathbb{Z}_T \rtimes \mathbb{Z}_2$ instead of a direct product. For $T=2$, we have $\sigma^{-1}=\sigma$, so that the dihedrality condition becomes the commutation of $\sigma$ and $\tau$.} $D_4$. Using this dihedrality condition and the facts that the Takiff currents $\mathcal{J}_{r,[p]}$ are valued in the real form $\mathfrak{g}$ and the positions $z_r$ and levels $\ell_{r,p}$ are real numbers, one checks that the twist function \eqref{Eq:Twist} and the Gaudin Lax matrix \eqref{Eq:Gaudin} satisfy the reality conditions
\begin{equation*}
\tau(\Gamma(z,x)) = \Gamma(\bar{z},x) \hspace{30pt} \text{ and } \hspace{30pt} \overline{\varphi(z)} = \varphi(\bar{z})\,,
\end{equation*}
which can be seen as equivariance conditions under the aforementioned action of $\mathbb{Z}_2$. Combining these with the conditions \eqref{Eq:Equivariance}, we then get that $\Gamma(z,x)$ and $\vp(z)$ are equivariant under the action of the full dihedral group $D_4$, as expected from the general construction of dihedral AGM in~\cite{Vicedo:2017cge}.

\subsection{Hamiltonian, constraint and gauge symmetry}
\label{Sec:Hamiltonian}

\paragraph{Zeroes of the twist function.} Let us begin by studying the zeroes of the twist function \eqref{Eq:Twist}. Firstly, we note that $z=0$ is always a zero of $\vp(z)$. We will suppose that this zero is simple, \textit{i.e.} that $\vp'(0)\neq 0$. Moreover, the behaviour of $\vp(z)$ at $z=\infty$ is described by the following asymptotic expansion:
\begin{equation}\label{Eq:AsymptTwist}
\vp\left( \frac{1}{u} \right) = 2K u^3 + O(u^5)\,, \qquad \text{ where } \qquad K = \frac{1}{2} \sum_{r=1}^N z_r \left(z_r \, \ell_{r,0} + 2 \, \ell_{r,1}\right)\,.
\end{equation}
Let us make a few comments on this expansion. From the equivariance property \eqref{Eq:Equivariance} of $\vp(z)$, it is clear that only odd powers of $u$ can appear in the expansion of $\vp(u^{-1})$ around $u=0$. Moreover, in general, the function $\vp(z)$ as defined in equation \eqref{Eq:Twist} also possesses a term of order $O(u)$ in its expansion at infinity, which is proportional to the sum $\sum_{r=1}^N \ell_{r,0}$: as we supposed that this sum vanishes (see the first-class condition \eqref{Eq:SumLevels}), the first term in the expansion is then of order $u^3$. Let us now consider the 1-form $\vp(z)\dd z$. To study its behaviour at infinity, let us consider the change of coordinate $z=u^{-1}$. We then have
\begin{equation} \label{Eq:Chi}
\vp(z) \dd z = \chi(u) \dd u\,, \qquad \text{ with } \qquad \chi(u) = -\frac{1}{u^2} \vp\left( \frac{1}{u} \right)\,.
\end{equation}
According to the asymptotic expansion \eqref{Eq:AsymptTwist}, the 1-form $\vp(z)\dd z$ thus has a zero at infinity. Moreover, the derivative of this 1-form at $z=\infty$ is given by $\chi'(0)=-2K$. We will suppose that this zero at infinity is simple, \textit{i.e.} that $K\neq 0$.

As $\vp(z)\dd z$ possesses $4N$ poles (counted with multiplicities), it possesses $4N-2$ zeroes in the Riemann sphere: in addition to the one at the origin $z=0$ and the one at infinity $z=\infty$, it thus possesses $4(N-1)$ zeroes in $\C\setminus\lbrace0\rbrace$. From the equivariance property \eqref{Eq:Equivariance}, one sees that these zeroes come as pairs $\zeta_i$ and $\zeta_{-i}=-\ze_i$, with $i\in\lbrace 1,\cdots,2N-2\rbrace$. We will suppose that the $\zeta_i$'s are pair-wise distinct and are thus simple zeroes of $\vp(z)$, hence $\vp'(\ze_i)\neq 0$. In terms of the $z_r$'s and the $\ze_i$'s, the twist function can then be rewritten as
\begin{equation}\label{Eq:TwistZeroes}
\varphi(z) = 2K\frac{z\prod_{i=1}^{2N-2}(z^2-\zeta_i^2)}{\prod_{r=1}^{N}(z^2-z_r^2)^2}\,.
\end{equation}

\paragraph{Hamiltonian.} Let us consider the following quantity:
\begin{equation}\label{Eq:Qz}
\mathcal{Q}(z) = -\frac{1}{2\varphi(z)} \int_{\mathbb{D}} \text{d}x \ \kappa(\Gamma(z,x),\Gamma(z,x))\,,
\end{equation}
which depends on the spectral parameter $z \in \mathbb{C}$. We define:
\begin{subequations}\label{Eq:Qi}
\begin{equation}
\mathcal{Q}_{\pm i} = \underset{z = \pm\zeta_i}{\operatorname{res}} \mathcal{Q}(z) \text{d}z, \hspace{15pt} i = 1,\ldots,2N-2\,,
\end{equation}
\begin{equation}
\mathcal{Q}_0 = \underset{z = 0}{\operatorname{res}} \; \mathcal{Q}(z) \text{d}z \hspace{30pt} \text{ and } \hspace{30pt} \mathcal{Q}_\infty = \underset{z = \infty}{\operatorname{res}} \mathcal{Q}(z) \text{d}z\,. 
\end{equation}
\end{subequations}
These quantities are local charges quadratic in the currents $\mathcal{J}_{r,[p]}$. It is straightforward to show that $\mathcal{Q}_i = \mathcal{Q}_{-i}$, from the equivariance property \eqref{Eq:Equivariance} of the Gaudin Lax matrix and twist function. Moreover, from \eqref{Eq:PoissonG}, one can prove that they are also in involution \textit{i.e.} they mutually Poisson commute. Given a collection of real numbers $\{\epsilon_0, \epsilon_i, \epsilon_\infty\}, \, i = 1,\ldots,2N-2$, we define the \textit{naive Hamiltonian} of the model (the term naive will be explained later in this section) as the following sum over the charges introduced above:
\begin{equation} \label{Hamiltonian}
\mathcal{H} = \epsilon_0 \mathcal{Q}_0 + 2\sum_{i=1}^{2N-2} \epsilon_i \mathcal{Q}_i + \epsilon_\infty \mathcal{Q}_\infty\,,
\end{equation}
where we introduce the factor of $2$ for future convenience. Due to the reality conditions introduced in the previous subsection, $\mathcal{H}$ can be shown to be real.

\paragraph{Constraint.} In this paragraph, we introduce a constraint on the phase space of canonical fields on $T^*G^N$ and show its consistency with the choice of Hamiltonian made in the previous paragraph. We will use the Dirac theory of constraints in Hamiltonian systems: we refer to~\cite{Dirac:1964,Henneaux:1992} for reviews of this formalism. Following the general construction of~\cite{Vicedo:2017cge}, we define the constraint as
\begin{equation}\label{Eq:Const1}
\Cc(x) = - \res_{z=\infty} \Gamma(z,x)\dd z = \lim_{u\to 0} \frac{1}{u}\Gamma\left( \frac{1}{u}, x\right).
\end{equation}
Using the expression \eqref{Eq:Gaudin} of the Gaudin Lax matrix $\Gamma(z,x)$ and the fact that $\frac{1}{2}(\Id+\s)$ is the projector on the grading $\g^{(0)}$ of $\g$, one checks that the constraint explicitly reads
\begin{equation}\label{Eq:Const2}
\Cc(x) = \sum_{r=1}^N \mathcal{J}_{r,[0]}^{(0)}(x)\,.
\end{equation}
In particular, it is a $\g^{(0)}$-valued field. The models we are interested in are then defined on a reduced phase space, obtained from canonical fields on $T^*G^N$ by imposing
\begin{equation*}
\Cc(x) \approx 0\,.
\end{equation*}
In this equation, and in the rest of this article, we use the notation $\approx$ to denote \textit{weak equalities}, \textit{i.e.} equalities that are true when the constraint is imposed. The standard equality sign $=$ will then indicate \textit{strong equalities}, which are true even without imposing the constraint.

\paragraph{Poisson bracket of the constraint with the naive Hamiltonian.} From the Poisson bracket \eqref{Eq:PoissonG} of the Gaudin Lax matrix with itself, one checks that the local charge $\Q(z)$, defined in equation \eqref{Eq:Qz}, satisfies the following Poisson bracket with the constraint:
\begin{equation*}
\lbrace \Q(z), \Cc(x) \rbrace = -\p_x \Gamma(z,x)^{(0)}\,.
\end{equation*}
In particular, as $\Gamma(z,x)$ is regular at $z=0$ and $z=\ze_i$ for $i=1,\cdots,2N-2$, one has
\begin{equation}
\lbrace \Q_i, \Cc(x) \rbrace = 0, \qquad \forall\, i\in\lbrace 0,\cdots,2N-2 \rbrace\,,
\end{equation}
for the charges $\Q_i$ introduced in equation \eqref{Eq:Qi}. Moreover, the residue of $\Gamma(z,x)^{(0)}\dd z$ at $z=\infty$ is equal to $-\Cc(x)$. Thus we also have
\begin{equation}
\lbrace \Q_\infty, \Cc(x) \rbrace = \p_x \Cc(x)\,.
\end{equation}
Recall that the naive Hamiltonian of the model $\Hc$ is defined in terms of the charges $\Q_i$, $i\in\lbrace 0,\cdots,2N-2,\infty\rbrace$, by equation \eqref{Hamiltonian}. Thus, we get
\begin{equation*}
\lbrace \Hc, \Cc(x) \rbrace = \epsilon_\infty\p_x \Cc(x)\,.
\end{equation*}
In particular, we see that the naive Hamiltonian weakly Poisson commutes with $\Cc(x)$:
\begin{equation}\label{Eq:PBHC}
\lbrace \Hc, \Cc(x) \rbrace \approx 0\,.
\end{equation}
This ensures that the Hamiltonian flow of $\Hc$ preserves the constraint $\Cc(x) \approx 0$.

\paragraph{First-class property.} \label{Sec:First-class} The Poisson bracket of the constraint with itself can be obtained from its definition \eqref{Eq:Const1} and the Poisson bracket \eqref{Eq:PoissonG} of the Gaudin Lax matrix (or equivalently from its expression \eqref{Eq:Const2} and the Poisson bracket \eqref{Eq:PoissonJ} of the currents $\mathcal{J}_{r,[0]}$). It reads
\begin{equation}\label{Eq:PBCC}
\bigl\lbrace \Cc\ti{1}(x), \Cc\ti{2}(y) \bigr\rbrace = \bigl[ C^{(00)}\ti{12}, \Cc\ti{1}(x) \bigr] \delta_{xy}\,,
\end{equation}
where $C^{(00)}\ti{12}\in\g^{(0)}\otimes\g^{(0)}$ is the split Casimir of $\g^{(0)}$. In fact, this bracket also contains in general a non-ultralocal term $-\left(\sum_{r=1}^N \ell_{r,0}\right)C^{(00)}\ti{12}\delta'_{xy}$: as we supposed in equation \eqref{Eq:SumLevels} that the levels $\ell_{r,0}$ sum to zero, this term vanishes. In particular, this shows that the Poisson  bracket of the constraint with itself weakly vanishes:
\begin{equation}\label{Eq:FirstClass}
\bigl\lbrace \Cc\ti{1}(x), \Cc\ti{2}(y) \bigr\rbrace \approx 0\,.
\end{equation}
Thus, the constraint $\Cc(x)\approx 0$ is \textit{first-class} (see for instance~\cite{Dirac:1964,Henneaux:1992}). This justifies a posteriori the name of first-class condition for the assumption \eqref{Eq:SumLevels} that we made: indeed, without this assumption, the bracket of the constraint would contain a non-ultralocal term which would not vanish weakly and the constraint would then not be first-class.

\paragraph{Total Hamiltonian and Lagrange multiplier.} \label{Sec:TotalHamiltonian} In the beginning of this subsection, we defined the naive Hamiltonian $\Hc$ through equation \eqref{Hamiltonian}. As we are considering models subject to the constraint $\Cc(x) \approx 0$, we have to define the \textit{total Hamiltonian} of the system as the sum of the naive Hamiltonian and a generic term proportional to the constraint, so that it coincides weakly with the naive Hamiltonian. It thus takes the form
\begin{equation}\label{Eq:HamTot}
\Hc_T = \Hc + \int_\D dx\;\kappa\bigl( \mu(x),\Cc(x) \bigr)\,,
\end{equation}
where $\mu$ is a $\g^{(0)}$-valued field, called the \textit{Lagrange multiplier}. It is a new dynamical field, independent of the canonical fields on $T^*G^N$. As we shall see in the next paragraph, the existence of this Lagrange multiplier reflects the presence of a gauge symmetry in the model.\\

The dynamic of the model is defined by the Hamiltonian flow of $\Hc_T$, \textit{i.e.} the time evolution of any observable $\Oc$ is given by
\begin{equation}\label{Eq:Dyn}
\p_t \Oc \approx \lbrace \Hc_T, \Oc \rbrace \approx \lbrace \Hc, \Oc \rbrace + \int_\D dx\;\kappa\bigl( \mu(x), \lbrace \Cc(x), \Oc \rbrace \bigr)\,.
\end{equation}
The facts that the naive Hamiltonian Poisson commutes with the constraint (see equation \eqref{Eq:PBHC}) and that the constraint is first-class (see equation \eqref{Eq:FirstClass}) ensure that the constraint $\Cc(x)\approx 0$ is conserved under time evolution:
\begin{equation}
\p_t \Cc(x) \approx 0\,.
\end{equation}

\paragraph{Gauge symmetry.} \label{Sec:Gauge} It is a standard result that the presence of first-class constraints in Hamiltonian systems implies the existence of gauge (local) symmetries (see for instance~\cite{Dirac:1964,Henneaux:1992}). The infinitesimal action of this gauge symmetry on the observables of the model is given by the Hamiltonian flow generated by the constraint. In the case at hand, the constraint satisfies the bracket \eqref{Eq:PBCC}, which is a copy of the Kirillov-Kostant bracket of the Lie algebra $\g^{(0)}$ for every point $x\in\D$. Thus, the gauge symmetry takes the form of a local action of the group $G^{(0)}$. The corresponding infinitesimal transformation of an observable $\Oc$, with gauge parameter $\epsilon(x,t) \in \g^{(0)}$, is given by
\begin{equation}\label{Eq:Gauge}
\delta_\epsilon \Oc \approx \left\lbrace \int_{\D} dx \; \kappa\bigl( \epsilon(x,t), \Cc(x) \bigr), \Oc \right\rbrace  \approx \int_{\D} dx \; \kappa\bigl( \epsilon(x,t), \left\lbrace \Cc(x), \Oc \right\rbrace \bigr)\,.
\end{equation}
One can then observe that the terms involving the Lagrange multiplier $\mu$ in the total Hamiltonian \eqref{Eq:HamTot} and the dynamic \eqref{Eq:Dyn} of the model correspond to a gauge transformation and thus account for the freedom of performing such a transformation in the time evolution of the system.\\

Let us study in more details the action of the gauge symmetry on the canonical fields on $T^*G^N$. For that, recall the expression \eqref{Eq:Const2} of the constraint $\Cc(x)$ in terms of the Kac-Moody currents $\mathcal{J}_{r,[0]}$. It is clear from the definition \eqref{Eq:KM} of the latter and the Poisson brackets \eqref{Eq:PoissongX}, \eqref{Eq:Poissonj} and \eqref{Eq:PoissonW1} that
\begin{equation} \label{Eq:PoissonCg}
\bigl\lbrace \Cc\ti{1}(x), g_r\null\ti{2}(y) \bigr\rbrace = g_r\null\ti{2}(x)C\ti{12}^{(00)}\delta_{xy}\,.
\end{equation}
Thus, using equation \eqref{Eq:Gauge}, one finds that the infinitesimal gauge transformation of the field $g_r(x)$ is given by:
\begin{equation}\label{Eq:GaugeG}
\delta_\epsilon g_r(x) = g_r(x) \epsilon(x,t)\,.
\end{equation}
Similarly, one can determine the gauge transformation of the fields $X_r$. It is in fact more convenient to consider the gauge transformation of the field $Y_r=X_r+\ell_{r,0} W_r/2$, which reads
\begin{equation}\label{Eq:GaugeX}
\delta_\epsilon Y_r(x) = \left[Y_r(x), \epsilon(x,t) \right] + \frac{\ell_{r,0}}{2} \p_x \epsilon(x,t)\,.
\end{equation}
The transformations \eqref{Eq:GaugeG} and \eqref{Eq:GaugeX} are infinitesimal actions with local parameter $\epsilon(x,t)$ valued in $\g^{(0)}$. They can be lifted to an action of the group $G^{(0)}$, depending on a local parameter $h(x,t)$ in $G^{(0)}$, which takes the form:
\begin{equation}\label{Eq:GaugeCan}
g_r \longmapsto g_r h \qquad \text{ and } \qquad Y_r \longmapsto h^{-1}Y_r h + \frac{\ell_{r,0}}{2}\, h^{-1}\p_x h\,.
\end{equation}
In particular, we see that the gauge symmetry acts on the set of fields $(g_1,\cdots,g_N)\in G^N$ by right translation of the diagonal subgroup
\begin{equation*}
G^{(0)}_{\diag} = \bigl\lbrace (h,\cdots,h), \; h\in G^{(0)} \bigr\rbrace\,.
\end{equation*}

Let us summarise what are the physical degrees of freedom of the model. By construction, we start from the phase space of canonical fields on $T^*G^N$. One then needs to restrict to the field configurations such that the constraint $\Cc(x) \approx 0$ is satisfied. Furthermore, one needs to quotient out by the action of the gauge symmetry \eqref{Eq:GaugeCan} (the fact that this gauge symmetry preserves the constraint $\Cc(x) \approx 0$ is a direct consequence of the first-class property \eqref{Eq:FirstClass} of $\Cc(x)$). As explained above, this gauge symmetry acts on the coordinate fields $(g_1,\cdots,g_N)\in G^N$ by right translation of the subgroup $G^{(0)}_{\diag}$: one can then see the ``physical'' coordinate fields of the model as fields on the quotient $G^N/G^{(0)}_{\diag}$, by gauging away the coordinate fields in $G^{(0)}_{\diag}$. The constraint $\Cc(x)\approx 0$ can then be seen as eliminating the corresponding superfluous conjugate momentum fields. The physical phase space of the model can thus be identified with canonical fields on $T^*(G^N/G^{(0)}_{\diag})$: in particular, the Lagrangian formulation of the model will then describe a field theory on $G^N/G^{(0)}_{\diag}$. In this article, we will however keep working with the unreduced phase space $T^*G^N$, together with the constraint and the gauge symmetry, to avoid having to consider the quotient.

\paragraph{Gauge transformation of the Gaudin Lax matrix.} From the expression \eqref{Eq:Const2} of the constraint and the Poisson bracket \eqref{Eq:PoissonJ} of the Takiff currents $\mathcal{J}_{r,[p]}(x)$, one checks that the gauge transformation of the latter is given by
\begin{equation}
\delta_\epsilon \mathcal{J}_{r,[p]}(x) = [\mathcal{J}_{r,[p]}(x),\epsilon(x,t)] + \ell_{r,p}\, \p_x \epsilon(x,t)\,.
\end{equation}
The corresponding lifted action of the group $G^{(0)}$, with local parameter $h(x,t)\in G^{(0)}$, reads
\begin{equation} \label{Eq:GaugeJ}
\mathcal{J}_{r,[p]} \longmapsto h^{-1} \mathcal{J}_{r,[p]} h + \ell_{r,p}\;h^{-1}\p_x h\,.
\end{equation}
Recall that $\sigma$ is an automorphism of $G$ whose fixed-points form the subagroup $G^{(0)}$ and which induces an automorphism of $\g$ that leaves the elements of the subalgebra $\g^{(0)}$ invariant. As $h\in G^{(0)}$ and $h^{-1}\p_x h \in \g^{(0)}$, the gauge transformation of $\sigma\bigl( \mathcal{J}_{r,[p]}\bigr)$ is given by
\begin{equation*}
\sigma\bigl( \mathcal{J}_{r,[p]} \bigr) \longmapsto  h^{-1} \sigma\bigl( \mathcal{J}_{r,[p]} \bigr) h + \ell_{r,p}\;h^{-1}\p_x h\,.
\end{equation*}
From equations \eqref{Eq:Twist} and \eqref{Eq:Gaudin}, we then see that the gauge symmetry acts on the Gaudin Lax matrix $\Gamma(z)$ as
\begin{equation} \label{Eq:GaugeGaudinL}
\Gamma(z) \longmapsto  h^{-1} \Gamma(z) h + \vp(z)\;h^{-1}\p_x h\,.
\end{equation}

\paragraph{Gauge transformation of the Lagrange multiplier.} It is a standard result that the equations of motion of the model are invariant under gauge symmetries, as one should expect, if one also transforms the Lagrange multiplier appropriately~\cite{Dirac:1964,Henneaux:1992}. In the present case, the transformation rule of the Lagrange multiplier is
\begin{equation}
\delta_\epsilon \mu(x) = \bigl[ \mu(x),\epsilon(x,t) \bigr] + \p_t \epsilon(x,t) - \epsilon_\infty \p_x \epsilon(x,t) \,. 
\end{equation}
This infinitesimal transformation can be lifted to the following action of the gauge group $G^{(0)}$, with local parameter $h(x,t)\in G^{(0)}$:
\begin{equation}\label{Eq:GaugeMu}
\mu \longmapsto h^{-1}\mu h + h^{-1}\p _t h - \epsilon_\infty\, h^{-1}\p_x h\,.
\end{equation}

\subsection{Space-time symmetries} \label{Sec:Momentum}

In this section, we discuss the space-time symmetries of the models constructed in this article and in particular find a simple condition for their relativistic invariance.

\paragraph{Momentum.} Recall that the momentum of the phase space consisting of canonical fields on $T^*G$ is given by equation \eqref{Eq:Momentum}. The model constructed in the previous subsections is defined on $N$ copies of this phase space and thus has the following momentum:
\begin{equation}
\Pc = \sum_{r=1}^N \Pc_r = \sum_{r=1}^N \int_{\D} \dd x\; \kappa\bigl( X_r(x), j_r(x) \bigr)\,.
\end{equation}
Using the fact that
\begin{subequations}
\begin{equation}
\Gamma(z) = \frac{1}{2} \frac{\ell_{r,1} j_r}{(z-z_r)^2} + \frac{1}{2} \frac{X_r+\ell_{r,0}W_r/2+\ell_{r,0}j_r/2}{z-z_r} + O\bigl( (z-z_r)^0 \bigr)\,, \vspace{-4pt}
\end{equation}
\begin{equation}
\frac{1}{\vp(z)} = 2(z-z_r)^2 \left( \frac{1}{\ell_{r,1}} - \frac{\ell_{r,0}}{\ell_{r,1}^2}(z-z_r) + O\bigl( (z-z_r)^2 \bigr) \right),
\end{equation}
\end{subequations}
together with the definition \eqref{Eq:Qz} of $\Q(z)$ and the orthogonality relation \eqref{Eq:Orthogonal}, one checks that
\begin{equation}
\res_{z=z_r} \Q(z)\dd z = -\frac{1}{2} \Pc_r \,.
\end{equation}
Similarly, one finds that the residue at $-z_r$ gives the exact same result. Thus, one has
\begin{equation}
\Pc=-\sum_{r=1}^N \left( \res_{z=z_r} \Q(z)\dd z + \res_{z=-z_r} \Q(z)\dd z \right).
\end{equation}
Recall from subsection \ref{Sec:Hamiltonian} that, in addition to its poles at the sites $z_r$ and their opposites $-z_r$, $\Q(z)\dd z$ has poles at the zeroes of the twist function $0, \ze_1, -\ze_1, \cdots, \ze_{2N-2}, -\ze_{2N-2}, \infty$, with corresponding residues $\Q_0, \Q_1, \Q_1, \cdots, \Q_{2N-2}, \Q_{2N-2}, \Q_\infty$ (in particular the residues of $Q(z)\dd z$ at $\ze_i$ and $-\ze_i$ are equal for $i\in\lbrace 1,\cdots,2N-2\rbrace$). Combining the above expression of $\Pc$ with the fact that the sum of the residues of $\Q(z)\dd z$ vanishes, we finally get
\begin{equation}\label{Eq:MomQi}
\Pc = \Q_0 + 2 \sum_{i=1}^{2N-2} \Q_i + \Q_\infty\,.
\end{equation}
This shows that the momentum of the model possesses a simple expression in terms of the charges $\Q_i$. Let us note the similarity of this expression with the one \eqref{Hamiltonian} of the naive Hamiltonian $\Hc$: one sees that the momentum would correspond to the choice of all coefficients $\epsilon_i$ equal to $1$ in equation \eqref{Hamiltonian}. This allows a parallel treatment of the conserved charges associated to spatial translation (momentum) and temporal translation (Hamiltonian).

\paragraph{Energy-momentum tensor.} To explore further the space-time symmetries of the model, let us describe its energy-momentum tensor $\T\mu\nu$ (we will use greek labels $\mu$ and $\nu$ to describe space-time components, with $\mu,\nu=0$ corresponding to temporal components and $\mu,\nu=1$ to spatial ones). The components $\T00$ and $\T01$ are defined as the densities of respectively the Hamiltonian and the momentum of the model\footnote{As we are discussing conserved charges, we can work with weak equalities and thus define $\T00$ from the naive Hamiltonian and not the total one.}:
\begin{equation}
\Hc = \int_\D \dd x \; \T00(x) \qquad \text{ and } \qquad \Pc = \int_\D \dd x \; \T01(x)\,.
\end{equation}
Let us introduce, for $i\in\lbrace 0,\cdots,2N-2,\infty\rbrace$,
\begin{equation}
q_i(x) = \res_{z=\ze_i} q(z,x)\dd z\,, \qquad \text{ where } \qquad q(z,x) = -\frac{1}{2\varphi(z)} \kappa(\Gamma(z,x),\Gamma(z,x))\,.
\end{equation}
The fields $q_i(x)$ are then the densities of the local charges $\Q_i = \int_\D \dd x\;q_i(x)$. From the expressions \eqref{Hamiltonian} and \eqref{Eq:MomQi} of the Hamiltonian and momentum of the model, we then get
\begin{equation}\label{Eq:T0}
\T00 = \epsilon_0\, q_0 + 2\sum_{i=1}^{2N-2} \epsilon_i\, q_i + \epsilon_\infty\, q_\infty \qquad \text{ and } \qquad \T01 = q_0 + 2\sum_{i=1}^{2N-2} q_i + q_\infty\,.
\end{equation}
The other two components $\T10$ and $\T11$ of the energy-momentum tensor are defined through the local conservation law $\p_\mu \T\mu\nu=0$ obeyed by $\T\mu\nu$ as a consequence of the space-time translation invariance of the model. Decomposing this conservation equation in components, we get
\begin{equation}\label{Eq:Cons}
\p_t \T0\mu + \p_x \T1\mu =0\,, \qquad \text{ for } \mu=0,1\,.
\end{equation}
In order to find the expression of $\T1\mu$, we thus need to determine the time evolution of $\T0\mu$. A direct computation starting from the bracket \eqref{Eq:PoissonG} shows that the densities $q_i$ defined above satisfy
\begin{equation*}
\bigl\lbrace q_i(x), q_j(y) \bigr\rbrace \approx -\delta_{ij} \lambda_i \bigl( \p_x q_i(x)\delta_{xy} + 2 q_i(x) \delta'_{xy} \bigr)\,, ~~~ \text{ where } ~~~ \lambda_i = \left\lbrace \begin{array}{ll}
1 & \text{ if } i=0,\infty\,, \\
1/2 & \text{ if } i=1,\cdots,2N-2\,.
\end{array} \right.
\end{equation*}
From this equation, one easily deduces the evolution of $q_i(x)$ under the Hamiltonian flow of $\Q_j$, namely $\lbrace \Q_j, q_i(x) \rbrace \approx \delta_{ij}\lambda_i\, \p_x q_i(x)$. To obtain the time evolution of $q_i(x)$, one needs to take into account the Lagrange multiplier term in the dynamics \eqref{Eq:Dyn}. One shows that this term in fact does not contribute, as the densities $q_i(x)$ are first-class and more precisely satisfy $\lbrace \Cc(y), q_i(x) \rbrace = \delta_{i\infty} \Cc(x) \delta'_{xy} \approx 0$. Thus, the time evolution of $q_i(x)$ is given by (note that the factor $2$ in front of the charges $\Q_i$ in the Hamiltonian for $i\in\lbrace 1,\cdots,2N-2\rbrace$ cancels with the factor $\lambda_i$)
\begin{equation}
\p_t q_i = \epsilon_i \, \p_x q_i\,.
\end{equation}
Using the expressions \eqref{Eq:T0} of $\T00$ and $\T01$, we get
\begin{align*}
\p_t \T00 &= \epsilon_0^2\, \p_x q_0 + 2\sum_{i=1}^{2N-2} \epsilon_i^2\, \p_x q_i + \epsilon_\infty^2\, \p_x q_\infty\,, \\
\p_t \T01 &= \epsilon_0\, \p_x q_0 + 2\sum_{i=1}^{2N-2} \epsilon_i\, \p_x q_i + \epsilon_\infty\, \p_x q_\infty\,.
\end{align*}
Comparing to the conservation equation \eqref{Eq:Cons}, we get the components $\T10$ and $\T11$ of the energy momentum tensor:
\begin{equation}\label{Eq:T1}
\T10 = -\epsilon_0^2\, q_0 - 2\sum_{i=1}^{2N-2} \epsilon_i^2\, q_i - \epsilon^2_\infty\, q_\infty \;\;\; \text{ and } \;\;\; \T11 = -\epsilon_0\, q_0 - 2\sum_{i=1}^{2N-2} \epsilon_i\, q_i - \epsilon_\infty q_\infty\,.
\end{equation}

\paragraph{Classical scale invariance.} From equations \eqref{Eq:T0} and \eqref{Eq:T1}, we note that $\T\mu\mu=\T00+\T11=0$. The energy momentum tensor is thus traceless. It is a standard result in field theory that this implies the classical scale invariance of the model. In general, this scale invariance is expected to be broken at the quantum level. However, as we shall see in subsection \ref{SubSec:Limit}, some particular limit of the model that we are constructing will also maintain this scale invariance at the quantum level and define a conformal field theory.

\paragraph{Relativistic invariance.} Let us introduce the two-dimensional Minkowski metric $\eta_{\mu\nu}$, defined by $\eta_{00}=-\eta_{11}=1$ and $\eta_{01}=\eta_{10}=0$, and the tensor $T_{\mu\nu}=\eta_{\mu\rho}\T\rho\nu$ obtained by lowering one of the index of the energy-momentum tensor. From equations \eqref{Eq:T0} and \eqref{Eq:T1}, we get
\begin{equation}
T_{01} = q_0 + 2\sum_{i=1}^{2N-2} q_i + q_\infty \qquad \text{ and } \qquad T_{10} = \epsilon_0^2\, q_0 + 2\sum_{i=1}^{2N-2} \epsilon_i^2\, q_i + \epsilon^2_\infty\, q_\infty\,.
\end{equation}
It is a standard result of field theory that the model is invariant under Lorentz symmetries (preserving the metric $\eta_{\mu\nu}$) if the tensor $T_{\mu\nu}$ is symmetric and thus if $T_{01}=T_{10}$. It is clear from the above equation that this is the case if and only if the coefficients $\epsilon_i$ all square to $1$, \textit{i.e.}
\begin{equation}\label{Eq:EpsRel}
\epsilon_i = \pm 1\,, \qquad \forall \, i\in\lbrace 0,\cdots,2N-2,\infty \rbrace\,.
\end{equation}
This gives a particularly simple condition ensuring the relativistic invariance of the model\footnote{Let us briefly discuss the converse of this result. In general, the sufficient and necessary condition for relativistic invariance of the model is that the energy-momentum tensor is symmetric up to a total derivative. In the present case, this is equivalent to $T_{01}-T_{10}=(1-\epsilon_0^2) q_0 + 2\sum_{i=1}^{2N-2} (1-\epsilon_i^2)q_i + (1-\epsilon_\infty^2)q_\infty$ being a total derivative. From the definition of the densities $q_i$, there is no apparent choice of $\epsilon_i$'s which would make this combination a total derivative, expect for taking all coefficients $1-\epsilon_i^2$ equal to 0. Thus, we expect the condition \eqref{Eq:EpsRel} to also be a necessary condition for the relativistic invariance of the model.}.

\subsection{Integrability}
\label{SubSec:Integrability}

\paragraph{Lax matrix.} Following \cite{Vicedo:2017cge}, we define the Lax matrix of the model as the following $\mathfrak{g}^{\mathbb{C}}$-valued field:
\begin{equation} \label{Eq:LaxDef}
\mathcal{L}(z,x) = \frac{\Gamma(z,x)}{\varphi(z)}\,.
\end{equation}
To give an explicit description of this Lax matrix, let us determine its partial fraction decomposition. As $\Gamma(z)$ and $\vp(z)$ have the same poles (at the points $z_i$ and $-z_i$), of the same order, $\Lc(z)$ has poles at the zeroes of the twist function $\vp(z)$, \textit{i.e.} at $z=0$, $z=\pm\ze_i$ for $i\in\lbrace 1,\cdots,2N-1 \rbrace$ and $z=\infty$. One easily checks that the residues of $\Lc(z)$ at $z=0$ and $\ze_i$ are respectively equal to $\Gamma(0)/\vp'(0)$ and $\Gamma(\ze_i)/\vp'(\ze_i)$. Moreover, using the equivariance properties \eqref{Eq:Equivariance}, one finds that the residue of $\Lc(z)$ at $z=-\ze_i$ is equal to $\Gamma(-\ze_i)/\vp'(-\ze_i)=-\sigma\bigl(\Gamma(\ze_i)\bigr)/\vp'(\ze_i)$. This fixes the non-polynomial part of the partial fraction decomposition of $\Lc(z)$. To determine the polynomial part, let us study the behaviour of $\Lc(z)$ around $z=\infty$. The asymptotic expansion of the Gaudin Lax matrix $\Gamma(z,x)$ around infinity reads
\begin{equation}\label{Eq:AsymptGamma}
\Gamma\left(\frac{1}{u},x\right) = u\, \mathcal{C}(x) - u^2 \mathcal{B}(x) - u^3 \mathcal{B}_1(x) + O(u^4) \approx - u^2 \mathcal{B}(x) - u^3 \mathcal{B}_1(x) + O(u^4)\,,
\end{equation}
where $\mathcal{B}(x)$ and $\mathcal{B}_1(x)$ are the following $\mathfrak{g}$-valued currents:
\begin{subequations} \label{Eq:B}
\begin{align}
\mathcal{B}(x) &= -\sum_{r=1}^N \left( z_r \mathcal{J}_{r,[0]}^{(1)} + \mathcal{J}_{r,[1]}^{(1)}\right), \\
\mathcal{B}_1(x) &= -\sum_{r=1}^N z_r \left(z_r\mathcal{J}_{r,[0]}^{(0)} + 2 \mathcal{J}_{r,[1]}^{(0)}\right).
\end{align}
\end{subequations}
Moreover, using the expression \eqref{Eq:TwistZeroes} of the twist function, we get
\begin{equation}\label{Eq:AsymptInversePhi}
\frac{1}{\vp(1/u)} = \frac{1}{u^3}\left( \frac{1}{2K} + O(u^2) \right). 
\end{equation}
Using the asymptotic expansions \eqref{Eq:AsymptGamma} and \eqref{Eq:AsymptInversePhi}, one can then express the $O(u^{-1})$ and $O(u^0)$-terms in the expansion of $\Lc(1/u)$ around $u=0$, which correspond to the linear and constant terms in the polynomial part of the partial fraction decomposition of $\Lc(z)$. In the end, we then get
\begin{equation} \label{Eq:L}
\mathcal{L}(z,x) \approx \frac{1}{\varphi'(0)} \frac{\Gamma(0,x)}{z} + \sum_{i=1}^{2N-2}\sum_{k=0}^1 \frac{1}{\varphi'(\zeta_i)} \frac{(-1)^{k}\sigma^k\bigl(\Gamma(\zeta_i,x)\bigr)}{z-(-1)^{k}\zeta_i}  - \frac{\mathcal{B}_1(x)}{2K} - \frac{\mathcal{B}(x)}{2K}z\,.
\end{equation}

\paragraph{Lax connection and zero curvature equation.} Together with another $\mathfrak{g}^{\mathbb{C}}$-valued field $\mathcal{M}(z,x)$, $\Lc(z,x)$ forms the Lax connection of the model, \textit{i.e.} the equations of motion of the model can be recast as the zero curvature equation
\begin{equation}\label{Eq:ZCEq}
\partial_t\mathcal{L}(z,x) - \partial_x\mathcal{M}(z,x) + [\mathcal{M}(z,x),\mathcal{L}(z,x)] = 0\,.
\end{equation}
This was proven for general affine Gaudin models in~\cite{Vicedo:2017cge}. Let us briefly re-derive it in the present case and show the explicit expression of $\Mc(z,x)$. For that, we have to study the dynamic of the Lax matrix $\Lc(z)$, which is induced by the Hamiltonian flow \eqref{Eq:Dyn} of the total Hamiltonian $\Hc_T$. From the Poisson bracket \eqref{Eq:PoissonG}, one finds that the Poisson bracket of the charge $\Q(w)$ (defined in equation \eqref{Eq:Qz}) with the Lax matrix $\Lc(z,x)$ is given by
\begin{equation}\label{Eq:PBQwL}
\bigl\lbrace \Q(w), \Lc(z,x) \bigr\rbrace = \bigl[ \Lc(z,x), \Mc(w \, ;z,x) \bigr] + \p_x \Mc(w \, ;z,x) - \p_x \left( \frac{1}{\vp(z)} \kappa\ti{2}\Bigl(\Rc^0\ti{21}(w,z), \Gamma\ti{2}(w,x) \Bigr) \right),
\end{equation}
where we defined
\begin{equation}
\Mc(w \, ;z,x) = -\frac{1}{\vp(w)} \kappa\ti{2}\Bigl(\Rc^0\ti{12}(z,w), \Gamma\ti{2}(w,x) \Bigr)\,.
\end{equation}
Let us note that the Hamiltonian flow \eqref{Eq:PBQwL} induced by $\Q(w)$ on $\Lc(z,x)$ almost takes the form of a zero curvature equation, up to the last term. To deduce the Hamiltonian flow induced by the charges $\Q_i$, $i\in\lbrace 0,\cdots,2N-2,\infty\rbrace$, defining the Hamiltonian, one has to take residues of the bracket \eqref{Eq:PBQwL} at $w=\ze_i$, where for uniformity we introduce the notation $\ze_0=0$ and $\ze_\infty=\infty$. Let us note that $\Rc\ti{21}(w,z)$ and $\Gamma(w,x)$ are regular at $w=\ze_i$ if $\ze_i$ is finite, \textit{i.e.} if $i\in\lbrace 0,1,\cdots,2N-2\rbrace$: thus, in this case, the last term in the bracket \eqref{Eq:PBQwL} does not possess a residue at $w=\ze_i$. A similar statement holds for the residue at infinity: $\Gamma(1/u,x)$ and $\Rc\ti{21}(1/u,z)$ are both of order $O(u)$ around $u=0$, so that the last term in the bracket \eqref{Eq:PBQwL} for $w=1/u$ is of order $O(u^2)$ and thus defines a regular 1-form at $w=\infty$. We then get that the Hamiltonian flow of $\Q_i$, $i\in\lbrace 0,\cdots,2N-2,\infty \rbrace$, on $\Lc(z,x)$ takes the form of a zero curvature equation:
\begin{equation*}
\lbrace \Q_i, \Lc(z,x) \rbrace - \p_x \Mc_i(z,x) + \bigl[ \Mc_i(z,x), \Lc(z,x) \bigr] = 0\,, \quad \text{ with } \quad \Mc_i(z,x) = \res_{w=\ze_i} \Mc(w;z,x)\dd w\,.
\end{equation*}
For $i\in\lbrace 0,\cdots,2N-2 \rbrace$, a direct computation gives
\begin{equation*}
\Mc_i(z,x) = \frac{1}{2\vp'(\ze_i)} \sum_{k=0}^1 \frac{(-1)^k\sigma^k \bigl( \Gamma(\ze_i,x) \bigr)}{z-(-1)^{k}\ze_i}\,.
\end{equation*}
From the equivariance property \eqref{Eq:Equivariance}, one finds that $\sigma\bigl(\Gamma(0,x)\bigr)=-\Gamma(0,x)$. Thus, we get in particular that $\Mc_0(z,x)=\Gamma(0)/z\vp'(0)$. To compute $\Mc_\infty(z,x)$, we use the asymptotic expansions \eqref{Eq:AsymptGamma} and \eqref{Eq:AsymptInversePhi}, as well as
\begin{equation}\label{Eq:AsymptR}
\Rc\ti{12}^0 \left( z, \frac{1}{u} \right) = u \, C\ti{12}^{(00)} + u^2 z\,C\ti{12}^{(11)} + O(u^3)\,.
\end{equation}
After a short computation, we get:
\begin{equation}
\Mc_\infty(z,x) \approx - \frac{\mathcal{B}_1(x)+z\,\mathcal{B}(x)}{2K}\,.
\end{equation}
To complete the derivation of the temporal part $\Mc(z,x)$ of the Lax connection, we finally need to compute the contribution of the Lagrange multiplier $\mu$ to the dynamics of $\Lc(z,x)$. From the Poisson bracket \eqref{Eq:PoissonG}, the definition \eqref{Eq:Const1} of the constraint and the expansion \eqref{Eq:AsymptR} we get
\begin{equation}
\bigl\lbrace \Cc\ti{2}(y), \Lc\ti{1}(z,x) \bigr\rbrace = -\bigl[ C\ti{12}^{(00)}, \Lc\ti{1}(z,x) \bigr] \delta_{xy} + C\ti{12}^{(00)}\,\delta'_{xy}\,.
\end{equation}
Thus,
\begin{equation*}
\int_\D dy\;\kappa\ti{2}\bigl( \mu\ti{2}(y), \lbrace \Cc\ti{2}(y), \Lc\ti{1}(z,x) \rbrace \bigr) = - \bigl[ \mu(x), \Lc(z,x) \bigr] + \p_x\mu(x)\,.
\end{equation*}
Combining all the results above, we find that the dynamics of $\Lc(z,x)$ follows the zero curvature equation \eqref{Eq:ZCEq}, for $\Mc(z,x)$ given by
\begin{equation}\label{Eq:M}
\mathcal{M}(z,x) \approx \frac{\epsilon_0}{\varphi'(0)} \frac{\Gamma(0,x)}{z} + \sum_{i=1}^{2N-2}\sum_{k=0}^1 \frac{\epsilon_i}{\varphi'(\zeta_i)} \frac{(-1)^{k}\sigma^k \bigl(\Gamma(\zeta_i,x)\bigr)}{z-(-1)^{k}\zeta_i} - \epsilon_\infty\frac{\mathcal{B}_1(x)}{2K} - \epsilon_\infty\frac{\mathcal{B}(x)}{2K}z + \mu(x)\,.
\end{equation}

\paragraph{Maillet bracket and integrability.} Since the equations of motion of the model take the form of a zero curvature equation, one can extract an infinite number of conserved charges from the monodromy of the Lax matrix $\mathcal{L}(z, x)$. The integrability of the model is a consequence of the fact that these charges are in involution. In order to show this, one starts from the bracket of the Lax matrix. In our case it is simply computed from the Poisson bracket \eqref{Eq:PoissonG} of the Gaudin Lax matrix with itself and reads:
\begin{multline} \label{PoissonL}
 \ \ \ \ \ \ \ \ \{\mathcal{L}_{\underline{\mathbf{1}}}(z,x),\mathcal{L}_{\underline{\mathbf{2}}}(w,y)\} = [\mathcal{R}_{\underline{\mathbf{12}}}(z,w),\mathcal{L}_{\underline{\mathbf{1}}}(z,x)] \delta_{xy} - [\mathcal{R}_{\underline{\mathbf{21}}}(w,z),\mathcal{L}_{\underline{\mathbf{2}}}(w,x)] \delta_{xy} \\ 
- (\mathcal{R}_{\underline{\mathbf{12}}}(z,w) + \mathcal{R}_{\underline{\mathbf{21}}}(w,z)) \delta'_{xy}\,, \ \ \ \ \ \ \ \ \  \ \ \ \ \ \ \ \  \ \ \ \ \ \ \ \  \ \ \ \ \ \ \ \  
\end{multline}
where $\mathcal{R}_{\underline{\mathbf{12}}}(z,w) = \mathcal{R}^0_{\underline{\mathbf{12}}}(z,w)\varphi(w)^{-1}$ and $\mathcal{R}^0$ is the twisted standard $\mathcal{R}$-matrix \eqref{Eq:DefRMatrix}.
The bracket \eqref{PoissonL} is an example of a Maillet non-ultralocal bracket \cite{Maillet:1985fn,Maillet:1985ek}. It satisfies the Jacobi identity due to the fact that the $\mathcal{R}$-matrix is a solution of the classical Yang-Baxter equation:
\begin{equation*}
[\mathcal{R}_{\underline{\mathbf{12}}}(z_1,z_2),\mathcal{R}_{\underline{\mathbf{13}}}(z_1,z_3)] + [\mathcal{R}_{\underline{\mathbf{12}}}(z_1,z_2),\mathcal{R}_{\underline{\mathbf{23}}}(z_2,z_3)] + [\mathcal{R}_{\underline{\mathbf{32}}}(z_3,z_2),\mathcal{R}_{\underline{\mathbf{13}}}(z_1,z_3)] = 0\,,
\end{equation*}
which is a consequence of the fact that $\mathcal{R}^0$ is also a solution (see equation \eqref{Eq:CYBE}).
It is a standard result that the Maillet bracket implies the involution of the charges extracted from the monodromy of the Lax matrix $\mathcal{L}(z, x)$.

\paragraph{Integrable local hierarchies.} Let us consider the charges $\Q_i$, $i\in\lbrace 0,\cdots,2N-2,\infty \rbrace$, defined in equation \eqref{Eq:Qi}. For $i\neq \infty$, a direct computation shows that
\begin{equation}\label{Eq:Qi2}
\Q_i = -\frac{1}{2\vp'(\ze_i)} \int_\D \dd x \; \kappa\bigl( \Gamma(\ze_i,x), \Gamma(\ze_i,x) \bigr)\,. 
\end{equation}
Similarly, one shows that the charge $\Q_\infty$ admits the following weak expression:
\begin{equation}\label{Eq:QInf}
\Q_\infty \approx -\frac{1}{2\chi'(0)} \int_\D \dd x \; \kappa\bigl( \mathcal{B}(x), \mathcal{B}(x) \bigr)\,.
\end{equation}
The function $\chi(u)$ was introduced in equation \eqref{Eq:Chi} to describe the 1-form $\vp(z)\dd z$ around infinity, while the field $\mathcal{B}(x)$ can be seen as the evaluation of the 1-form $\Gamma(z,x)\dd z$ at $z=\infty$. The above expression is then a natural generalisation for $i=\infty$ of equation \eqref{Eq:Qi2}.

The quadratic charges $\Q_i$ are naturally associated with the zeroes of the twist function. In fact, it was shown in~\cite{Lacroix:2017isl} that in addition to the non-local charges extracted from the monodromy, models with twist function admit infinite towers of local conserved charges in involution, obtained from the zeroes of the twist function and which generalise the construction of the quadratic charges $\Q_i$. These towers, which are called integrable hierarchies, consist of charges of increasing degrees whose densities are well-chosen invariant polynomials\footnote{The degrees of these polynomials follow a specific pattern, which depends on the underlying Lie algebra $\g$ and the zero which is considered. For zeroes $\ze_i$, $i\in\lbrace 1,\cdots,2N-2 \rbrace$, which are not fixed under the $\mathbb{Z}_2$-transformation $z\mapsto-z$, these degrees are given by one plus the exponents of the untwisted affine algebra of $\g$. For the zeroes $0$ and $\infty$, which are fixed under $z\mapsto -z$, only a subset of the exponents appears, which depends on the choice of automorphism $\s$ (see~\cite{Lacroix:2017isl} for more details).} of $\Gamma(\ze_i,x)$ for $i\neq \infty$ and of $\mathcal{B}(x)$ for $i=\infty$ (only weakly in this case). The charge of lowest degree in each tower is quadratic and the corresponding invariant polynomial is simply the bilinear form $\kappa(\cdot,\cdot)$, thus giving back the charges $\Q_i$ considered above. We refer to~\cite{Lacroix:2017isl} for more details about the construction of these hierarchies (for completeness, let us note that these local charges were first constructed in~\cite{Evans:1999mj,Evans:2000hx} for the Principal Chiral Model, with and without Wess-Zumino term, and in~\cite{Evans:2000qx} for the symmetric coset $\s$-model, which corresponds to the model considered here for $N=1$).

\paragraph{Lax connection in light-cone coordinates.} For completeness, let us briefly comment on the expression of the Lax connection in light-cone components. Let us consider the light-cone coordinates $x^\pm = (t \pm x)/2$ and the corresponding derivatives $\partial_\pm = \partial_t \pm \partial_x$. The zero curvature equation \eqref{Eq:ZCEq} can then be rewritten as
\begin{equation*}
\partial_+\mathcal{L}_-(z) - \partial_-\mathcal{L}_+(z) + [\mathcal{L}_+(z),\mathcal{L}_-(z)] = 0\,,
\end{equation*}
where we have introduced the light-cone Lax connection
\begin{equation*}
\mathcal{L}_\pm(z) = \mathcal{M}(z) \pm \mathcal{L}(z)\,.
\end{equation*}
From the expressions \eqref{Eq:L} and \eqref{Eq:M} of $\mathcal{L}(z)$ and $\mathcal{M}(z)$ respectively, we observe that they contain the same terms. For the case of $\mathcal{M}(z)$, these terms are multiplied by one of the coefficients $\epsilon_i$, $i\in\lbrace 0,\cdots,2N-2,\infty \rbrace$, which we recall can be either $+1$ or $-1$ to ensure the relativistic invariance of the model. Hence, depending on the values of these numbers, these terms will be present only in one of the two light-cone components of the Lax connection. More precisely, they will appear in the expression of $\mathcal{L}_+(z)$ if the corresponding $\epsilon_i$ is equal to $+1$ and in the expression of $\mathcal{L}_-(z)$ if $\epsilon_i$ is equal to $-1$. This determines the pole structure of these two quantities.

\paragraph{Gauge symmetry and integrable structure.} Let us now discuss how the integrable structure of the model behaves under the $G^{(0)}_{\text{diag}}$ gauge symmetry introduced in subsection \ref{Sec:Gauge} and in particular determine how the Lax connection transforms under this symmetry. From its definition \eqref{Eq:LaxDef} and the transformation \eqref{Eq:GaugeGaudinL} of the Gaudin Lax matrix, one simply finds that $\mathcal{L}(z)$ transforms as
\begin{equation} \label{Eq:GaugeL}
\mathcal{L}(z) \longmapsto  h^{-1} \mathcal{L}(z) h + h^{-1}\p_x h\,.
\end{equation}
Let us now focus on $\mathcal{M}(z)$. From \eqref{Eq:GaugeGaudinL}, we obtain that the evaluations of the Gaudin Lax matrix at finite zeros of the twist function vary covariantly as $\Gamma(0) \mapsto  h^{-1} \Gamma(0) h$ and $\Gamma(\zeta_i) \mapsto  h^{-1} \Gamma(\zeta_i) h$. Moreover, inserting the asymptotic expansions \eqref{Eq:AsymptGamma} and \eqref{Eq:AsymptTwist} in equation \eqref{Eq:GaugeGaudinL}, we get
\begin{equation*}
\mathcal{B} \longmapsto h^{-1} \mathcal{B} h \qquad \text{ and } \qquad \mathcal{B}_1 \longmapsto h^{-1} \mathcal{B}_1 h -2K \,h^{-1}\p_x h\,.
\end{equation*}
Combining the above results with the transformation \eqref{Eq:GaugeMu} of the Lagrange multiplier $\mu$ and the expression \eqref{Eq:M} of $\mathcal{M}(z)$, one then finds that $\mathcal{M}(z)$ transforms as
\begin{equation} \label{Eq:GaugeM}
\mathcal{M}(z) \longmapsto  h^{-1} \mathcal{M}(z) h + h^{-1}\p_t h\,.
\end{equation}
Re-expressing the equations \eqref{Eq:GaugeL} and \eqref{Eq:GaugeM} in light-cone components, we finally arrive at
\begin{equation}\label{Eq:GaugeLC}
\mathcal{L}_\pm(z) \longmapsto  h^{-1} \mathcal{L}_\pm(z) h + h^{-1}\p_\pm h\,.
\end{equation}

Let us make a few comments. Firstly, we note that the transformation \eqref{Eq:GaugeLC} takes the form of a formal gauge transformation $\mathcal{L}_\pm(z) \mapsto \mathcal{L}_\pm^h(z)$ of the Lax connection. Such formal gauge transformations are a general feature of integrable field theories, regardless of whether they possess a gauge symmetry or not. They can be performed for any $h = h(z,x,t)$ in the group $G$ and leave the zero curvature equation invariant: they therefore encode the non-uniqueness of the Lax connection in the integrable field theory under consideration. In the present case, this field theory also possesses a $G^{(0)}_{\text{diag}}$ gauge symmetry, which encodes the presence of unphysical degrees of freedom in the model. The above computation thus shows that the action of this gauge symmetry, with local parameter $h(x,t)\in G^{(0)}$, on the Lax connection coincides with a formal gauge transformation with parameter $h$. Since such a transformation preserves the zero curvature equation, which is a reformulation of the equations of motion of the model, this provides an alternative check of the invariance of these equations of motion under the $G^{(0)}_{\text{diag}}$ gauge symmetry.

Moreover, it is a standard result that the conserved charges extracted from the monodromy of the Lax matrix are invariant under formal gauge transformations. As a consequence, the above results show that these charges are also invariant with respect to the $G^{(0)}_{\text{diag}}$ gauge symmetry of the model.\\

Recall that in addition to these charges extracted from the monodromy matrix, the model also admits an infinite number of local conserved charges in involution (see above). The latter are also gauge invariant, as was proven in general in~\cite{Lacroix:2017isl}. This fact can also be checked directly using the results derived above. Indeed, we have shown that the currents $\Gamma(0,x)$, $\Gamma(\ze_i,x)$ and $\mathcal{B}(x)$ are covariant under gauge transformations. As mentioned earlier in this subsection, the densities of the local conserved charges are obtained by taking conjugacy invariant polynomials of these currents, which are then gauge invariant.

\subsection{The panorama of the models} \label{Sec:Panorama}

Let us end this section by briefly discussing the panorama of integrable models constructed above. A model in this class first depends on the number of sites $N$ of the underlying AGM, which fixes its target space $G^N/G^{(0)}_{\text{diag}}$. Moreover, following the different steps of the construction of the model, one sees that it is characterised by the following parameters:\vspace{-2pt}
\begin{itemize}\setlength\itemsep{0.1em}
\item the positions $z_1,\cdots,z_N$ of the sites ;
\item the levels $\ell_{1,0},\cdots,\ell_{N,0}$ and $\ell_{1,1},\cdots,\ell_{N,1}$ ;
\item the coefficients $\epsilon_0,\cdots,\epsilon_{2N-2},\epsilon_\infty$ entering the definition of the Hamiltonian \eqref{Hamiltonian}.
\end{itemize}
In particular, the parameters in the first two bullets are encoded in the twist function \eqref{Eq:Twist}.
As explained in subsection \ref{Sec:Momentum}, the coefficients $\epsilon_i$ cannot take arbitrary values as they are required to be either $+1$ or $-1$ to ensure the relativistic invariance of the model. Recall also that the levels $\ell_{r,0}$ are subject to the first-class condition \eqref{Eq:SumLevels}, which imposes one relation between them. Moreover, one shows that the model obtained by considering a dilation of the spectral parameter $z \mapsto az$ is equivalent to the inital model: this induces a redundancy among the parameters of the model, which can be fixed for instance by setting one of the position $z_r$ to a fixed value, say $z_1=1$. Thus, the model depends in the end on $3N-2$ continuous free parameters.\\

Recall from subsection \ref{Sec:Hamiltonian} that the definition of the Hamiltonian of the model involves the zeroes $\lbrace 0,\infty,\zeta_1,\cdots,\zeta_{2N-2} \rbrace$ of the twist function. In general, expressing these zeroes in terms of the positions $z_r$ and the levels $\ell_{r,p}$ is a complicated, if not impossible, task, as it requires solving a polynomial equation of degree $2N-2$. To circumvent this difficulty, one can choose another set of parameters of the model, given by:
\begin{itemize}\setlength\itemsep{0.1em}
\item the positions $z_2,\cdots,z_N$ of the sites (fixing $z_1=1$) ;
\item the constant term $K$ in the twist function \eqref{Eq:TwistZeroes} ;
\item the zeroes $\zeta_1,\cdots,\zeta_{2N-2}$ of the twist function and the corresponding coefficients $\epsilon_i\in \lbrace 
+1,-1 \rbrace$ ;
\item the coefficients $\epsilon_0$ and $\epsilon_\infty$ in $\lbrace 
+1,-1 \rbrace$.
\end{itemize}
This set of parameters is encoded in the choice of the twist function in its factorised form \eqref{Eq:TwistZeroes} (except for the discrete parameters $\epsilon_i=\pm 1$). In particular, if these are chosen as the defining parameters of the model, the levels $\ell_{r,p}$ are defined in terms of this expression of the twist function as the residues
\begin{equation*}
\ell_{r,0} = 2\underset{z = z_r}{\operatorname{res}} \vp(z)\text{d}z \hspace{30pt} \text{and} \hspace{30pt} \ell_{r,1} = 2\underset{z = z_r}{\operatorname{res}} (z-z_r)\vp(z)\text{d}z\,.
\end{equation*}
Note that in this parametrisation, the first-class condition \eqref{Eq:SumLevels} is automatically satisfied, as the factorised form \eqref{Eq:TwistZeroes} of the twist function ensures that $\vp(z)\dd z$ is regular at $z=\infty$. The $3N-2$ continuous parameters listed above are thus unconstrained.\\

Let us end this section by discussing briefly the simplest example in this panorama of models, the model with one site, \textit{i.e.} $N=1$. This model was first considered in~\cite{Vicedo:2017cge}, where it was shown that it coincides with the standard $\sigma$-model on the symmetric space $G/G^{(0)}$. In the parametrisation discussed above, this model possesses one site with fixed position $z_1=1$ and no zeroes $\ze_i$ ($0$ and $\infty$ are the only zeroes of the twist function). The only continuous free parameter of the model is then the constant term $K$. The twist function simply reads
\begin{equation}
\vp(z) = \frac{2K z}{(z^2-1)^2}\,.
\end{equation}
We fix the coefficients $\epsilon_i$ to\footnote{The choice  $\epsilon_\infty=-1$ and $\epsilon_0=+
1$ would simply lead to the opposite Hamiltonian, while the choices $\epsilon_\infty=\epsilon_0=\pm 1$ would lead to the Hamiltonian coinciding with (plus or minus) the momentum of the theory, as one can see from equation \eqref{Eq:MomQi}.} $\epsilon_\infty=+1$ and $\epsilon_0=-1$. The phase space of the model consists of canonical fields on a single copy of $T^*G$, described by the two fields $g(x)$ and $X(x)$ (as $N=1$, we drop the indices $r$). A direct computation shows that the naive Hamiltonian of the model \eqref{Hamiltonian} is given in this case by
\begin{equation}\label{Eq:NEq1}
\Hc_{N=1} = \frac{1}{2} \int_{\D} \dd x \; \left( \frac{1}{K} \kappa\bigl(X^{(1)},X^{(1)}\bigr) + K \kappa\bigl(j^{(1)},j^{(1)}\bigr) + 2 \kappa\bigl(X^{(0)},j^{(0)}\bigr) \right),
\end{equation}
where $j=g^{-1}\p_x g$ as above. As expected, this coincides with the Hamiltonian of the symmetric space $\sigma$-model on $G/G^{(0)}$, formulated as a model on $G$ with a $G^{(0)}$ gauge symmetry. In the present case, the constraint associated with this gauge symmetry simply reads $X^{(0)} \approx 0$.

\section{Lagrangian formulation of the models with two copies} \label{Sec:Model}

The Lagrangian formulation of the models we are concerned with in this article consists of field theories with fundamental fields $g_r(x,t)$, $r\in\lbrace 1,\cdots,N\rbrace$, taking values in $G$. We will obtain these Lagrangian theories by performing an inverse Legendre transform of the models constructed in section \ref{Sec:HamiltonianForm} in the Hamiltonian formulation. In order to make the computation of the inverse Legendre transform more explicit, we will restrict to the case of two copies, \textit{i.e.} we will fix $N = 2$.

Before that, let us briefly describe, as a simple illustration, the model with only one copy. The model is described in its Lagrangian formulation by a unique $G$-valued field $g(x,t)$. Performing the inverse Legendre transform of the Hamiltonian \eqref{Eq:NEq1}, one finds that its action takes the form:
\begin{equation*}
S_{N=1}[g] = \frac{K}{2} \iint_{\D\times\R} \dd x \, \dd t \; \kappa\bigl( j_+^{(1)}, j_-^{(1)} \bigr)\,,
\end{equation*}
where $j_\pm=g^{-1}\p_\pm g$. As expected, this is the action of the standard symmetric space $\sigma$-model on $G/G^{(0)}$ in its gauged formulation. One easily checks that this action is invariant under the gauge transformation $g(x,t)\mapsto g(x,t)h(x,t)$ for $h(x,t)\in G^{(0)}$.

Let us return to the models with $N=2$. Before proceeding to the computation of the inverse Legendre transform, let us describe the parameters of these models. From the discussion in subsection \ref{Sec:Panorama}, they depend on four continuous parameters: the position $z_2$ of the second site (having fixed the position of the first site to $z_1=1$), the global factor in the twist function $K$, and the zeroes $\zeta_1$ and $\zeta_2$. In the following we will rename $z_2 = x$ to avoid unnecessary indices,  although we will sometimes use the notation $z_1$ and $z_2$ so that some formulae assume a more compact form. In addition to these continuous parameters, the models are characterised by the choice of four discrete coefficients $(\epsilon_0, \epsilon_1, \epsilon_2, \epsilon_\infty)$ in $\{-1,+1\}$. 
We will fix these coefficients to the values\footnote{Other choices would give either equivalent models, up to a redefinition of the parameters, or models for which the inverse Legendre transform is singular and thus which do not possess a Lagrangian formulation.} $\epsilon_0 = -1,\epsilon_1 = -1,\epsilon_2 = +1$ and $\epsilon_\infty = +1$. Motivated by this choice and for future convenience, we will rename $\zeta_1$ as $\zeta_-$ and $\zeta_2$ as $\zeta_+$.

\subsection{Lagrangian expression of the momentum fields}

In order to perform the inverse Legendre transform of the models, we first need to express their momentum fields, encoded in the fields $X_r$ introduced in the previous section, in terms of the time derivatives of the coordinate fields $g_r$, encoded in the temporal Maurer-Cartan currents $j_{0,r} = g_r^{-1}\p_t g_r$.

For that, let us calculate the dynamics of the fields $g_r$, given by the Poisson bracket of $g_r$ with the total Hamiltonian introduced in subsection \ref{Sec:TotalHamiltonian}. We start by seeking a more explicit expression of the naive Hamiltonian \eqref{Hamiltonian} in terms of the fields $j_r$ and
\begin{equation*}
Y_r = X_r + \frac{\ell_{r,0}}{2} W_r\,,
\end{equation*}
which we introduce for future convenience. After a few manipulations, one rewrites it in the form
\begin{equation} \label{Eq:HamiltonianCoeff}
\mathcal{H} = \sum_{r,s=1}^2\sum_{k=0}^1 a_{rs}^{(k)} \int_{\mathbb{D}} \text{d}x \ \kappa\left(j_r^{(k)},j_s^{(k)}\right) + b_{rs}^{(k)} \int_{\mathbb{D}} \text{d}x \ \kappa\left(Y_r^{(k)},j_s^{(k)}\right) + c_{rs}^{(k)} \int_{\mathbb{D}} \text{d}x \ \kappa\left(Y_r^{(k)},Y_s^{(k)}\right),
\end{equation}
where the coefficients $a_{rs}^{(k)}, b_{rs}^{(k)}$ have slightly long expressions and are hence written in appendix \ref{Sec:Coefficients}, while the coefficients $c_{rs}^{(k)}$ are given by
\begin{equation} \label{Eq:Cs}
c_{rs}^{(0)} = \frac{\zeta_-^2}{2K} \, \frac{z_{\bar{r}}^2z_{\bar{s}}^2/\zeta_-^2 - z_{\bar{r}}^2 - z_{\bar{s}}^2 + \zeta_-^2}{\zeta_-^2 - \zeta_+^2} \qquad \text{ and } \qquad 
c_{rs}^{(1)} = \frac{z_r z_s}{2K} \, \frac{z_{\bar{r}}^2z_{\bar{s}}^2/\zeta_+^2 - z_{\bar{r}}^2 - z_{\bar{s}}^2+ \zeta_-^2}{\zeta_-^2 - \zeta_+^2}\,,
\end{equation}
where we introduced the notation $\bar{r} = 3 - r$ for $r = 1,2$.

The form \eqref{Eq:HamiltonianCoeff} of the Hamiltonian allows us to calculate easily what $j_{0,r}$ reads in terms of the Hamiltonian fields $j_r$ and $Y_r$. From the Poisson brackets \eqref{Eq:PoissongX}, \eqref{Eq:Poissonj} and \eqref{Eq:PoissonW1}, as well as the identity \eqref{Eq:CasGrad}, one shows that
\begin{equation*}
g_r^{-1}\{\mathcal{H},g_r\} = \sum_{s=1}^2\sum_{k=0}^1 b_{rs}^{(k)} j_s^{(k)} + 2 c_{rs}^{(k)} Y_s^{(k)}\,.
\end{equation*}
Hence, taking into account the form \eqref{Eq:HamTot} of the total Hamiltonian and the Poisson bracket \eqref{Eq:PoissonCg}, we have:
\begin{equation} \label{Eq:Lagrj0}
j_{0,r} \approx g_r^{-1}\{\mathcal{H}_T,g_r\} \approx \sum_{s=1}^2\sum_{k=0}^1 b_{rs}^{(k)} j_s^{(k)} + 2 c_{rs}^{(k)} Y_s^{(k)} + \mu\,.
\end{equation}
The above equation is a linear system that can be projected into the gradings and solved to express the fields $Y_r^{(k)}$ in terms of the currents $j_{0,r}$. However, we take a different path to eliminate the Lagrange multiplier $\mu$. For the grading zero, subtracting the equations for $r = 1$ and $r = 2$, we arrive at:
\begin{equation*}
2 \sum_{s=1}^2 \left(c_{1s}^{(0)} - c_{2s}^{(0)}\right) Y_s^{(0)} \approx j_{0,1}^{(0)} - j_{0,2}^{(0)} - \sum_{s=1}^2 \left(b_{1s}^{(0)} - b_{2s}^{(0)} \right) j_s^{(0)} \,.
\end{equation*}
In order to obtain a second equation independent of the Lagrange multiplier $\mu$, we make use of the constraint \eqref{Eq:Const2}, rewritten in the form:
\begin{equation*}
 Y_1^{(0)} + Y_2^{(0)} \approx - \frac{\ell_{1,0}}{2} j_1^{(0)} - \frac{\ell_{2,0}}{2} j_2^{(0)}\,.
\end{equation*}
Altogether, the solution for the grading zero is given by:
\begin{equation} \label{Eq:LagrY0}
Y_r^{(0)} \approx \frac{1}{2\sum_{s=1}^2 \left(c_{ss}^{(0)} - c_{s\bar{s}}^{(0)}\right)}\left(j_{0,r}^{(0)} - j_{0,\bar{r}}^{(0)} - \sum_{s=1}^2 \left(b_{rs}^{(0)} - b_{\bar{r}s}^{(0)} - \ell_{s,0} \left(c^{(0)}_{r\bar{r}} - c^{(0)}_{\bar{r}\bar{r}}\right)\right) j^{(0)}_s\right).
\end{equation}
For the grading one, one has the following equations:
\begin{equation*}
j_{0,r}^{(1)} \approx \sum_{s=1}^2 b_{rs}^{(1)} j_s^{(1)} + 2 c_{rs}^{(1)} Y_s^{(1)}\,.
\end{equation*}
If we rename the components of the inverse matrix of $(c^{(1)})_{rs} = c^{(1)}_{rs}$ as $\bar{c}^{(1)}_{rs} = (c^{(1)})^{-1}_{rs}$, the solution then reads:
\begin{equation} \label{Eq:LagrY1}
Y_r^{(1)} \approx \frac{1}{2}\sum_{s=1}^2\bar{c}_{rs}^{(1)}\left(j_{0,s}^{(1)} - \sum_{t=1}^2 b_{st}^{(1)} j_t^{(1)}\right).
\end{equation}

\subsection{Action of the model}

\paragraph{Inverse Legendre transform.} Using the definition of $X_r$ in terms of the canonical fields (see for instance \cite{Delduc:2019bcl} for more details), one shows that the action of the model is given by the following inverse Legendre transform\footnote{As we are now working in the Lagrangian formulation, in which the constraint always holds, we drop the distinction between weak and strong equalities. In particular, one can use the naive Hamiltonian (and not the total one) to compute the inverse Legendre transform.}:
\begin{equation*}
S[g_1,g_2] = \sum_{r=1}^2\iint\text{d}x \,\text{d}t \ \kappa\left(X_r,j_{0,r}\right) - \int \text{d}t \ \mathcal{H}\,.\end{equation*}
In terms of the fields $Y_r$ introduced in the previous subsection, we can rewrite the above equation as
\begin{equation*}
S[g_1,g_2] = \sum_{r=1}^2\iint\text{d}x \,\text{d}t \ \kappa\left(Y_r,j_{0,r}\right) - \int \text{d}t \ \mathcal{H} - \sum_{r=1}^2 \frac{\ell_{r,0}}{2}\, \Ww{g_r}\,,
\end{equation*}
where the Wess-Zumino terms of $g_r$ have now appeared, using equation \eqref{Wesszumino}. To obtain the explicit expression of the action, we now have to replace the Hamiltonian fields $Y_r$ by their Lagrangian expression, given by equations \eqref{Eq:LagrY0} and \eqref{Eq:LagrY1}, including in the Hamiltonian $\Hc$, using its expression \eqref{Eq:HamiltonianCoeff}. Let us introduce the light-cone components of the Maurer-Cartan currents $j_{\pm,r} = g_r^{-1}\p_\pm g_r = j_{0,r} \pm j_r$. After some manipulations, one finds
\begin{equation}\label{Eq:Action}
S[g_1,g_2] = \sum_{r,s=1}^2\iint\text{d}x \,\text{d}t \left( \rho_{rs}^{(0)}\,\kappa\left(j_{+,r}^{(0)},j_{-,s}^{(0)}\right) + \rho_{rs}^{(1)}\,\kappa\left(j_{+,r}^{(1)},j_{-,s}^{(1)}\right) \right) + \kay\, \Ww{g_1} - \kay\, \Ww{g_2}\,.
\end{equation}
In terms of the defining parameters of the model $K$, $x$, $\zeta_+$ and $\zeta_-$, the coefficients corresponding to the grading zero in this action are given by
\begin{subequations} \label{Eq:Coefficients}
\begin{equation}
\rho^{(0)}_{11} = \rho^{(0)}_{22} = \frac{K}{2} \frac{\zeta_-^2 - \zeta_+^2}{(1-x^2)^2}\,, \hspace{15pt} \rho^{(0)}_{12} = K\frac{\left(1-\zeta_+^2\right) \left(x^2-\zeta_-^2\right)}{\left(1-x^2\right)^3}\,, \hspace{15pt} \rho^{(0)}_{21} = -K\frac{\left(1-\zeta_-^2\right) \left(x^2-\zeta_+^2\right)}{\left(1-x^2\right)^3}\,,
\end{equation}
while the ones corresponding to the grading one are
\begin{align}
\rho^{(1)}_{11} &= \frac{K}{2}\frac{\left(1-2 \zeta_+^2+\zeta_-^2 \zeta_+^2\right)}{\left(1-x^2\right)^2}\,, \hspace{15pt} \rho^{(1)}_{12} = K\frac{x\left(1-\zeta_+^2\right) \left(x^2-\zeta_-^2\right)}{\left(1-x^2\right)^3}\,, \nonumber \\
\rho^{(1)}_{21} &= -K\frac{\left(1-\zeta_-^2\right) \left(x^2-\zeta_+^2\right)}{x \left(1-x^2\right)^3}\,, \hspace{15pt} \rho^{(1)}_{22} = \frac{K}{2}\frac{\left(x^4-2 \zeta_+^2 x^2+\zeta_-^2 \zeta_+^2\right)}{x^2\left(1-x^2\right)^2}\,.
\end{align}
Finally, the Wess-Zumino coefficient $\kay$ is defined as $\kay=-\ell_{1,0}/2 = \ell_{2,0}/2$ and explicitly reads
\begin{equation}
\kay = K\frac{2x^2 + 2\zeta_-^2\zeta_+^2 - (1 + x^2)(\zeta_-^2 + \zeta_+^2)}{(1 - x^2)^3}\,.
\end{equation}
\end{subequations}

\paragraph{Gauge symmetry.} Let us check explicitly that the action \eqref{Eq:Action} is invariant under the gauge transformation $g_r(x,t) \mapsto g_r(x,t) h(x,t)$ with $h(x,t) \in G^{(0)}$, as expected from the Hamiltonian construction. Under this transformation, the Wess-Zumino terms change according to the Polyakov-Wiegmann formula \cite{Polyakov:1983tt}:
\begin{equation*}
\Ww{g_r h} = \Ww{g_r} + \Ww{h} - \frac{1}{2} \iint \dd x \, \dd t \left[ \kappa\left(j_{+,r}^{(0)},(\p_-h) h^{-1}\right) - \kappa\left(j_{-,r}^{(0)},(\p_+h) h^{-1}\right) \right].
\end{equation*}
Moreover, the light-cone components of the Maurer-Cartan currents transform as:
\begin{align*}
j_{\pm,r}^{(0)} \longmapsto h^{-1}\bigl(j_{\pm,r}^{(0)} + (\p_\pm h)h^{-1} \bigr) h \qquad \text{ and } \qquad
j_{\pm,r}^{(1)} \longmapsto h^{-1} j_{\pm,r}^{(1)} h\,.
\end{align*}
It is then clear that in an action of the form \eqref{Eq:Action} with general coefficients $\rho_{rs}^{(k)}$ the terms of grading one are invariant under this gauge transformation. The variation of the action thus only contains terms in the grading zero, coming from the variation of the factors $\kappa\bigl(j_{+,r}^{(0)},j^{(0)}_{-,s}\bigr)$ and of the Wess-Zumino terms. Computing explicitly this variation, one finds that gauge invariance is verified if and only if the following conditions are satisfied:
\begin{equation} \label{Eq:GaugeCoefficients}
\rho_{11}^{(0)} + \rho_{12}^{(0)} - \frac{\kay}{2} = \rho_{12}^{(0)} + \rho_{22}^{(0)} - \frac{\kay}{2} = \rho_{21}^{(0)} + \rho_{22}^{(0)} + \frac{\kay}{2} = \rho_{11}^{(0)} + \rho_{21}^{(0)}  + \frac{\kay}{2} = 0\,.
\end{equation}
The above relations are indeed all identically satisfied for the choice of coefficients \eqref{Eq:Coefficients}.

Note that one can also rewrite the action \eqref{Eq:Action} in a manifestly gauge invariant way. Using the Polyakov-Wiegmann identity \cite{Polyakov:1983tt} to make the Wess-Zumino term $\Ww{g_1^{\null} g_2^{-1}}$ appear, as well as the relations \eqref{Eq:GaugeCoefficients}, one finds
\begin{eqnarray*}
S[g_1,g_2] &= & \displaystyle \rho_{11}^{(0)} \iint\text{d}x \,\text{d}t \;\kappa\left(j_{+,1}^{(0)}-j_{+,2}^{(0)},j_{-,1}^{(0)} - j_{-,2}^{(0)}\right) + \kay\, \Ww{g_1^{\null} g_2^{-1}} \\
& & \hspace{20pt} \displaystyle + \sum_{r,s=1}^2 \left( \rho_{rs}^{(1)} - \frac{\kay}{2}\epsilon_{rs} \right) \iint\text{d}x \;\text{d}t \,\kappa\left(j_{+,r}^{(1)},j_{-,s}^{(1)}\right) ,
\end{eqnarray*}
where $\epsilon_{12}=-\epsilon_{21}=1$ and $\epsilon_{11}=\epsilon_{22}=0$. As announced, this form of the action is manifestly invariant under a gauge transformation $g_r(x,t) \mapsto g_r(x,t) h(x,t)$ with $h(x,t) \in G^{(0)}$. Indeed, the field $g_1^{\null} g_2^{-1}$ is itself invariant and the currents $j_{\pm,1}^{(0)}-j_{\pm,2}^{(0)}$ and $j_{\pm,r}^{(1)}$ are covariant, \textit{i.e.} they transform as
\begin{align*}
j_{\pm,1}^{(0)}-j_{\pm,2}^{(0)} \longmapsto h^{-1}\bigl(j_{\pm,1}^{(0)}-j_{\pm,2}^{(0)} \bigr) h \qquad \text{ and } \qquad
j_{\pm,r}^{(1)} \longmapsto h^{-1} j_{\pm,r}^{(1)} h\,.
\end{align*}

\paragraph{Global symmetries.} \label{SubSec:GlobalSymm} Let us briefly discuss the global symmetries of the model \eqref{Eq:Action}, which are given by the left $(G\!\times\! G)$-translations on $g_1$ and $g_2$:
\begin{equation}
(g_1,g_2) \longmapsto (f_1g_1,f_2g_2), \qquad (f_1,f_2) \in G \times G\,.
\end{equation}
Indeed, these translations leave the Maurer-Cartan currents $j_{\pm,r}=g_r^{-1}\p_\pm g_r$ invariant and also preserve the Wess-Zumino terms $\Ww{g_r}$. Thus, they define global symmetries of the action \eqref{Eq:Action}. The conserved Noether currents associated to these symmetries read
\begin{subequations}
\begin{align*}
\mathcal{K}_{+,r} &= g_r \left( \sum_{s=1}^2 \sum_{k=0}^1 \rho_{sr}^{(k)} j_{+,s}^{(k)} - \frac{(-1)^{r}}{2} \kay \,j_{+,r} \right)g_r^{-1}\,, \\
\mathcal{K}_{-,r} &= g_r \left( \sum_{s=1}^2 \sum_{k=0}^1 \rho_{rs}^{(k)} j_{-,s}^{(k)} + \frac{(-1)^{r}}{2} \kay \,j_{-,r} \right)g_r^{-1}\,.
\end{align*}
\end{subequations}
These currents satisfy the conservation equation $\p_+ \mathcal{K}_{-,r} + \p_- \mathcal{K}_{+,r} = 0$. Let us also note that they are gauge-invariant under the $G^{(0)}_{\diag}$ gauge symmetry $g_r(x,t) \mapsto g_r(x,t) h(x,t)$ of the model.

\paragraph{Reformulation of the action.} As detailed in appendix \ref{App:Reformulation}, the coefficients $\rho_{rs}^{(k)}$ and $\kay_r$ (with $\kay_1=\kay$ and $\kay_2=-\kay$) defined in equation \eqref{Eq:Coefficients} can be re-expressed as residues of well-chosen functions (for the non-dihedral $\sigma$-models on $G^N$ defined in~\cite{Delduc:2018hty,Delduc:2019bcl}, a similar result was pointed out in~\cite{Lacroix:2019xeh}). This allows us to reformulate the action \eqref{Eq:Action} in the following remarkably simple way:
\begin{equation}\label{Eq:ActionRef1}
S = \sum_{r=1}^2 S_{\text{W}\hspace{-1pt}\text{Z}\hspace{-1pt}\text{W}\hspace{-1pt},\,\kay_r}[g_r] - 4K \iint \dd x \, \dd t\, \sum_{r,s=1}^2 \ \res_{w=z_s} \res_{z=z_r} \kappa\ti{12}\Bigl( \Rc^0\ti{12}(w,z)\vp_+(z)\vp_-(w), j_{+,r}\null\ti{1} \, j_{-,s}\null\ti{2} \Bigr)\,,
\end{equation}
where $\Rc^0\ti{12}$ is the $\Rc$-matrix \eqref{Eq:DefRMatrix} underlying the integrable structure of the model, $S_{\text{W}\hspace{-1pt}\text{Z}\hspace{-1pt}\text{W}\hspace{-1pt},\,\kay}[g]$ is the Wess-Zumino-Witten action
\begin{equation}\label{Eq:WZW}
S_{\text{W}\hspace{-1pt}\text{Z}\hspace{-1pt}\text{W}\hspace{-1pt},\,\kay}[g] = \frac{\kay}{2}\iint\text{d}x \,\text{d}t \ \kappa\left(g^{-1}\p_+g,g^{-1}\p_-g\right) + \kay\, \Ww{g}
\end{equation}
and $\vp_\pm(z)$ are functions defined as
\begin{equation}\label{Eq:PhiPM}
\vp_+(z) = \frac{z^2-\ze^2_+}{(z^2-z_1^2)(z^2-z_2^2)} \qquad \text{ and } \qquad \vp_-(z) = \frac{z(z^2-\ze^2_-)}{(z^2-z_1^2)(z^2-z_2^2)}\,.
\end{equation}
In particular, note that the reformulation \eqref{Eq:ActionRef1} of the action does not involve an explicit sum over the grading index $k=0,1$ as in the original expression \eqref{Eq:Action}. As explained in the appendix \ref{App:Reformulation}, this graded structure, and thus the choice of automorphism $\sigma$, is accounted for in the $\Rc$-matrix $\Rc^0\ti{12}$.

\paragraph{Conjectured generalisations.} Having derived equation \eqref{Eq:ActionRef1}, it is natural to formulate conjectures about generalisations of the models considered here. For instance, we expect a similar expression to hold for the models on $G^N/G^{(0)}_{\diag}$ with arbitrary $N$ constructed in the Hamiltonian formalism in section \ref{Sec:HamiltonianForm}. More generally, we conjecture that it also holds for models on $G^N/G^{(0)}_{\diag}$ with arbitrary $N$ and where the subalgebra $\g^{(0)}$ is the grading zero subspace of a $\mathbb{Z}_T$-gradation with arbitrary $T$, generalising the case $T=2$ considered here.

Let us be more precise about this conjecture. For $N=1$, the model on the $\mathbb{Z}_T$-coset $G/G^{(0)}$ for arbitrary $T$ was constructed in~\cite{Young:2005jv} and was identified with a realisation of $D_{2T}$-dihedral affine Gaudin model in~\cite{Vicedo:2017cge}, based on the Hamiltonian analysis carried out in~\cite{Ke:2011zzb}. Although the generalisations of this $\s$-model on cosets $G^N/G^{(0)}_{\diag}$ with arbitrary $N$ have not been considered before in the literature, we expect the procedure of section \ref{Sec:HamiltonianForm} to readily generalise to the construction of such models, using a $D_{2T}$-dihedral affine Gaudin model~\cite{Vicedo:2017cge} instead of a $D_4$-dihedral model. In this case, the twist function of the model would read\footnote{The equivariance condition \eqref{Eq:Equivariance} is then replaced by $\vp(\omega z) = \omega^{-1} \vp(z)$, where $\omega=\exp(2i\pi/T)$.}
\begin{equation}
\vp(z) = KT \frac{z^{T-1} \prod_{i=1}^{2N-2}(z^T-\ze_i^T)}{\prod_{r=1}^N (z^T-z_r^T)^2}\,,
\end{equation} 
in terms of its zeroes $\ze_1,\cdots,\ze_{2N-2}$ and poles $z_1,\cdots,z_N$. One can then factorise this twist function\footnote{As for the case $T=2$ treated in section \ref{Sec:HamiltonianForm}, we expect such a separation of the zeroes of $\vp(z)$ in two sets $\lbrace 0, \ze_1,\cdots,\ze_{N-1}\rbrace$ and $\lbrace \ze_N, \cdots,\ze_{2N-2}, \infty \rbrace$ to come naturally from the relativistic invariance of the model, which requires the coefficients $\epsilon_i$, $i\in\lbrace 0,1,\cdots,2N-2,\infty\rbrace$, in the Hamiltonian of the model to be  equal to either $-1$ or $+1$.} as $\vp(z)=KT \vp_+(z) \vp_-(z)$, similarly to equation \eqref{Eq:FactTwist} for $T=2$, with
\begin{equation*}
\vp_+(z) = \frac{\prod_{i=N}^{2N-2}\left(z^T-\ze^T_i\right)}{\prod_{r=1}^N\left(z^T-z_r^T\right)} \qquad \text{ and } \qquad \vp_-(z) = \frac{z^{T-1}\prod_{i=1}^{N-1}\left(z^T-\ze^T_i\right)}{\prod_{r=1}^N\left(z^T-z_r^T\right)}\,.
\end{equation*}
We then conjecture that the action of the model is given by
\begin{equation} \label{Eq:ActionRef}
S = \sum_{r=1}^N S_{\text{W}\hspace{-1pt}\text{Z}\hspace{-1pt}\text{W}\hspace{-1pt},\,\kay_r}[g_r] -  \frac{KT^3}{2} \iint \dd x \, \dd t\, \sum_{r,s=1}^N \ \res_{w=z_s} \res_{z=z_r} \kappa\ti{12}\Bigl( \Rc^0\ti{12}(w,z)\vp_+(z)\vp_-(w), j_{+,r}\null\ti{1} \, j_{-,s}\null\ti{2} \Bigr)\,,
\end{equation}
where $\kay_r= -\frac{T}{2} \res_{z=z_r} \vp(z)\dd z$ and $\Rc^0$ now denotes the $\mathbb{Z}_T$-graded $\Rc$-matrix which underlies the integrable structure of $D_{2T}$-dihedral affine Gaudin models~\cite{Vicedo:2017cge}, namely
\begin{equation*}
\Rc^0\ti{12}(w,z) = \sum_{k=0}^{T-1} \frac{w^kz^{T-1-k}}{z^T - w^T}\pi\ti{1}^{(k)}C\ti{12}\,,
\end{equation*}
with $\pi^{(k)}$, $k\in\lbrace 0,\cdots,T-1 \rbrace$, the projections along the grading $\g = \bigoplus_{k=0}^{T-1} \g^{(k)}$.

As mentioned above, for $N=1$ and arbitrary $T$, the corresponding integrable model on the $\mathbb{Z}_T$-coset $G/G^{(0)}$ has been constructed in~\cite{Young:2005jv}: we have checked that the action of this model can indeed be reformulated as in~\eqref{Eq:ActionRef}. Moreover, for the case of arbitrary $N$ and $T=1$, the results of~\cite{Lacroix:2019xeh} show that the action of the model is also given by \eqref{Eq:ActionRef}, with $\Rc^0\ti{12}(z,w)$ the standard non-twisted $\Rc$-matrix $C\ti{12}/(w-z)$. Finally, we have checked this conjecture by direct computation for all cases with $N \leq 3$ and $T \leq 3$.

\subsection{Lax connection in the Lagrangian formulation}

From the equations \eqref{Eq:L} and \eqref{Eq:M}, the Lax connection can be written in terms of the fields $j_r$, $Y_r$ and $\mu$. Moreover, from equation \eqref{Eq:Lagrj0}, we have:
\begin{equation*}
\mu \approx j_{0,r}^{(0)} - \sum_{s=1}^2 \left( b_{rs}^{(0)} j_s^{(0)} + 2 c_{rs}^{(0)} Y_s^{(0)} \right).
\end{equation*}
We can then express the Lax connection solely in terms of the fields $j_r$ and $Y_r$. Inserting equations \eqref{Eq:LagrY0} and \eqref{Eq:LagrY1}, we finally get the Lagrangian expression of the Lax connection. In terms of the light-cone currents $j_{\pm,r}$, it reads:
\begin{equation}\label{Eq:LaxLag}
\mathcal{L}_\pm(z) = \sum_{r=1}^2\sum_{k=0}^1\eta_{\pm,r}^{(k)}(z) j_{\pm,r}^{(k)}\,,
\end{equation}
where
\begin{align}\label{Eq:CoefficientsLax}
\eta_{\pm,1}^{(0)}(z) &= \frac{\left(z^2-x^2\right)\left(1-\zeta_\pm^2\right)}{\left(z^2-\zeta_\pm^2\right)\left(1-x^2\right)}\,, \hspace{30pt} \eta_{\pm,1}^{(1)}(z) = z^{\pm1}\,\eta_{\pm,1}^{(0)}(z)\,, \\
\eta_{\pm,2}^{(0)}(z) &= \frac{\left(z^2-1\right) \left(x^2-\zeta_\pm^2\right)}{\left(z^2-\zeta_\pm^2\right)\left(x^2-1\right)}\,, \hspace{30pt} \eta_{\pm,2}^{(1)}(z) = \left(\frac{z}{x}\right)^{\pm1}\,\eta_{\pm,2}^{(0)}(z)\,.
\end{align}
In particular, we note as an observation that $\eta_{\pm,s}^{(k)}(z_r) = \delta_{rs}$ (where we recall that $z_1=1$ and $z_2=x$) and therefore
\begin{equation}\label{Eq:LaxInterp}
\mathcal{L}_\pm(z_r) = j_{\pm,r}\,.
\end{equation}

\subsection{A limit of the model} \label{SubSec:Limit}

\paragraph{Definition of the limit.} Let us recall that the model with two copies introduced above depends on the four continuous real parameters $x$, $K$, $\zeta_+$ and $\zeta_-$. In this subsection, we will describe the simple form that this model assumes after taking a particular limit of these parameters. In particular, this limit will be our starting point in section \ref{Sec:T11}. We start by considering the following reparametrisation of $x$, $K$, $\zeta_+$ and $\zeta_-$ in terms of four new parameters $\alpha$, $\lambda_1$, $\lambda_2$ and $\lambda$:
\begin{equation} \label{Eq:Limit}
x = \frac{1}{\alpha}\,, \hspace{15pt} K = \frac{\lambda_2^2}{\alpha^2}\,,  \hspace{15pt} \zeta_+ = \frac{\lambda_1}{\lambda}\,, \hspace{15pt}  \zeta_- = \frac{\lambda}{\lambda_2\alpha}\,.
\end{equation}
We then define the limit we will be interested in by taking $\alpha \to 0$ while keeping the other parameters $\lambda_1$, $\lambda_2$ and $\lambda$ fixed.

\paragraph{Action.} Let us look at how the action of the model simplifies in this limit. From their expression \eqref{Eq:Coefficients}, we obtain that the coefficients $\rho_{rs}^{(k)}$ and $\kay$ simply become:
\begin{equation*}
\rho^{(0)}_{11} = \rho^{(0)}_{22} = \frac{\lambda^2}{2}\,, \hspace{15pt} \rho^{(0)}_{12} = \rho^{(1)}_{12} = \rho^{(1)}_{21} = 0\,, \hspace{15pt} \rho^{(0)}_{21} = -\kay = -\lambda^2\,, \hspace{15pt} \rho^{(1)}_{11} = \frac{\lambda_1^2}{2}\,, \hspace{15pt} \rho^{(1)}_{22} = \frac{\lambda_2^2}{2}\,.
\end{equation*}
Writing the action explicitly, we thus have
\begin{align}\label{Eq:LimitAction}
S[g_1,g_2] &= \iint\text{d}x \,\text{d}t\, \sum_{r=1}^2  \left( \frac{\lambda^2}{2}\,\kappa\left(j_{+,r}^{(0)},j_{-,r}^{(0)}\right) + \frac{\lambda_r^2}{2}\, \kappa\left(j_{+,r}^{(1)},j_{-,r}^{(1)}\right) \right) - \lambda^2\,\kappa\left(j_{+,2}^{(0)},j_{-,1}^{(0)}\right) \\[4pt]
&\hspace{35pt} +\lambda^2\, \Ww{g_1} - \lambda^2\, \Ww{g_2}\,. \notag
\end{align}

\paragraph{Lax connection.} \label{SubSec:LaxLimit} Let us now turn to the Lax connection. Taking the limit on the coefficients $\eta_{\pm,r}(z)$ defined in \eqref{Eq:CoefficientsLax} and reinserting in the expression \eqref{Eq:LaxLag} of the Lax connection, we get:
\begin{equation} \label{Eq:LimitLax1}
\mathcal{L}_+(z) = \frac{1}{\lambda^2z^2-\lambda_1^2}\left(\bigl(\lambda^2-\lambda_1^2\bigr) \left(j_{+,1}^{(0)} + z\, j_{+,1}^{(1)} \right) + \lambda^2\bigl(z^2-1\bigr)\, j_{+,2}^{(0)} \right), \hspace{15pt} \mathcal{L}_-(z) = j_{-,1}^{(0)} + \frac{j_{-,1}^{(1)}}{z}\,.
\end{equation}
One can check that the zero curvature equation for this Lax connection actually does not encode all the equations of motion of the model. To circumvent this difficulty, let us also consider the limit of $\Lc_\pm(z/\alpha)$, which we will denote as $\widetilde{\mathcal{L}}_\pm(z)$ (by construction, $\widetilde{\mathcal{L}}_\pm(z)$ also satisfies a zero curvature equation). A direct computation shows that 
\begin{equation} \label{Eq:LimitLax2}
\widetilde{\mathcal{L}}_+(z) = j_{+,2}^{(0)} + z\,j_{+,2}^{(1)}\,, \hspace{15pt} \widetilde{\mathcal{L}}_-(z) = \frac{1}{\lambda_2^2z^2-\lambda^2}\left(\lambda^2\bigl(z^2-1\bigr)\, j_{-,1}^{(0)} + \bigl(\lambda_2^2-\lambda^2\bigr) \left(z^2j_{-,2}^{(0)} + z\, j_{-,2}^{(1)} \right) \right).
\end{equation}
The combined zero curvature equations of $\mathcal{L}_\pm(z)$ and $\widetilde{\mathcal{L}}_\pm(z)$ are equivalent to all the equations of motion of the model.

\paragraph{Additional symmetry.} \label{SubSec:AdditionalSymm} For this paragraph, we will suppose that the pair $\bigl( G, G^{(0)} \bigr)$ characterising the model is such that $G^{(0)}$ possesses a center $Z$. There are many examples of such pairs, which include for instance $\bigl(SU(p+q),S\bigl(U(p)\times U(q)\bigr)\bigr)$, $\bigl(SL(p+q),S\bigl(GL(p)\times GL(q)\bigr)\bigr)$ and $\bigl( SO(2n), U(n) \bigr)$. As we will now show, in this case, the model \eqref{Eq:LimitAction} then possesses an additional global $Z$-symmetry, which acts on the fields $g_1,g_2 \in G$ as
\begin{equation}\label{Eq:AddSym}
(g_1,g_2) \longmapsto (g_1 k,g_2)\,, \qquad k\in Z\,.
\end{equation}
Note that we could also have considered the action $(g_1,g_2) \mapsto (g_1, g_2k)$, which is equivalent to the one above \textit{via} the $G^{(0)}_{\diag}$ gauge symmetry. Under the action \eqref{Eq:AddSym}, the graded components $j_{\pm,r}^{(k)}$ of the Maurer-Cartan currents transform as
\begin{equation}
j_{\pm,1}^{(0)} \longmapsto j_{\pm,1}^{(0)}\,, \qquad j_{\pm,1}^{(1)} \longmapsto k^{-1} j_{\pm,1}^{(1)} k\,, \qquad j_{\pm,2}^{(0)} \longmapsto j_{\pm,2}^{(0)} \qquad \text{ and } \qquad j_{\pm,2}^{(1)} \longmapsto j_{\pm,2}^{(1)}\,,
\end{equation}
where we have used the fact that $k$ is central in $G^{(0)}$ and thus that $k^{-1} j_{\pm,1}^{(0)} k= j_{\pm,1}^{(0)}$. Noting also that the Wess-Zumino term of $g_1$ is invariant under the transformation \eqref{Eq:AddSym}, \textit{i.e.} $\Ww{g_1k}=\Ww{g_1}$, it is direct to check that this transformation defines a symmetry of the action \eqref{Eq:LimitAction}, as claimed.

\paragraph{Guadagnini-Martellini-Mintchev model.}\label{Par:GMM} Let us now define $U = g_1$ and $V = g_2^{-1}$. We recall that the Wess-Zumino term satisfies the following relation:
\begin{equation*}
\Ww{g^{-1}} = -\Ww{g}\,.
\end{equation*}
Then, in the case in which $\lambda_1 = \lambda_2 = \lambda$, the action \eqref{Eq:LimitAction} can be rewritten as
\begin{equation} \label{Eq:ActionComp}
S[U,V] = S_{\text{W}\hspace{-1pt}\text{Z}\hspace{-1pt}\text{W}\hspace{-1pt},\,\lambda^2}[U] + S_{\text{W}\hspace{-1pt}\text{Z}\hspace{-1pt}\text{W}\hspace{-1pt},\,\lambda^2}[V] + \lambda^2\iint\text{d}x \,\text{d}t \ \kappa\left(\bigl(\p_+V V^{-1}\bigr)^{(0)},\bigl(U^{-1}\p_-U\bigr)^{(0)}\right)\,,
\end{equation}
where $S_{\text{W}\hspace{-1pt}\text{Z}\hspace{-1pt}\text{W}\hspace{-1pt},\,\kay}$ denotes the Wess-Zumino-Witten action with level $\kay$ as defined in \eqref{Eq:WZW}. The action \eqref{Eq:ActionComp} coincides with the one of the Guadagnini-Martellini-Mintchev model introduced in \cite{Guadagnini:1987ty} as a theory on $(G \times G')/H$, when considered in the special case $G' = G$ and $H = G^{(0)}$. This model was shown to preserve scale invariance at the quantum level at one loop in~\cite{Guadagnini:1987ty} and at two loops in~\cite{Belokurov:1990rp}. This thus shows that the integrable $\sigma$-model considered in this subsection is a two-dimensional conformal field theory for the specific choice $\lambda_1=\lambda_2=\lambda$ of its defining parameters. The Kac-Moody current algebras of this conformal model have been studied in~\cite{Guadagnini:1987qc}.

Let us finally note that in the case under consideration, the Lax connections $\mathcal{L}_\pm(z)$ and $\widetilde{\mathcal{L}}_\pm(z)$, given in \eqref{Eq:LimitLax1} and \eqref{Eq:LimitLax2} respectively, assume the following simple form:
\begin{align*}
\mathcal{L}_+(z) &= j_{+,2}^{(0)}\,, \hspace{15pt} \mathcal{L}_-(z) = j_{-,1}^{(0)} + \frac{j_{-,1}^{(1)}}{z}\,, \\
\widetilde{\mathcal{L}}_+(z) &= j_{+,2}^{(0)} + z\,j_{+,2}^{(1)}\,, \hspace{15pt} \widetilde{\mathcal{L}}_-(z) = j_{-,1}^{(0)}\,.
\end{align*}
The existence of a Lax connection for this model is consistent with the results of~\cite{Bardakci:1996gs}, where its integrability was first established.\footnote{The integrability of a class of models that includes \eqref{Eq:ActionComp} was also studied in~\cite{Georgiou:2016urf}.}

\section{Integrable \texorpdfstring{\bm{$\sigma$}}{sigma}-models on \texorpdfstring{\bm{$T^{1,1}$}}{T11} manifolds} \label{Sec:T11}

\subsection{The models}

\paragraph{Action.} Let us consider the model with two copies described in the previous section for the choice $G = SU(2)$, with Lie algebra $\g = \mathfrak{su}(2)$ generated by $I_a = i \sigma_a /2$, where $\sigma_a$ is the $a$-th Pauli matrix. We take $\sigma$ to be the $\mathbb{Z}_2$-automorphism of $\mathfrak{su}(2)$ defined by the following action on the generators: $\sigma(I_1) = -I_1$, $\sigma(I_2) = -I_2$ and $\sigma(I_3) = I_3$, so that $\g^{(0)} = \mathfrak{u}(1) = \text{span}\{I_3\}$ and correspondingly $G^{(0)} = U(1) = \exp(\R I_3)$. Let us finally pick the following parametrisation for the fields $(g_1,g_2) \in SU(2) \times SU(2)$ of the model:
\begin{subequations} \label{Eq:Parametrisation}
\begin{align}
g_1 &= \exp{(\phi_1I_3)}  \exp{(\theta_1 I_2)}  \exp{(\psi I_3)}\,, \\
g_2 &= \exp{(-\phi_2I_3)}  \exp{(-\theta_2 I_2)}  \exp{(-\widetilde{\psi} I_3)}\,.
\end{align}
\end{subequations}
Inserting this parametrisation in the action \eqref{Eq:LimitAction}, one finds:
{\small \begin{align} \label{Eq:ActionT11}
S &= \frac{1}{4}\hspace{-1pt} \iint \hspace{-1pt} \text{d}x \, \text{d}t \Bigl(\bigl(\lambda^2 \hspace{-1pt} + \hspace{-1pt} \lambda_1^2\hspace{-1pt} + \hspace{-1pt}\bigl(\lambda^2 \hspace{-1pt} - \hspace{-1pt} \lambda_1^2\bigr)\cos(2\theta_1)\bigr)\p_-\phi_1 \p_+\phi_1\hspace{-1pt} + \hspace{-1pt}2\lambda_1^2 \, \p_-\theta_1 \p_+\theta_1\hspace{-1pt} + \hspace{-1pt}2\lambda^2 \, \p_-\psi \p_+\psi\hspace{-1pt} + \hspace{-1pt}4\lambda^2 \cos \theta_1  \, \p_-\phi_1 \p_+\psi \nonumber \\
&\hspace{25pt} + \hspace{-1pt} \bigl(\lambda^2 \hspace{-1pt} + \hspace{-1pt} \lambda_2^2 \hspace{-1pt} + \hspace{-1pt} \bigl(\lambda^2 \hspace{-1pt} - \hspace{-1pt} \lambda_2^2\bigr)\cos(2\theta_2)\bigr)\p_-\phi_2 \p_+\phi_2 \hspace{-1pt} + \hspace{-1pt} 2\lambda_2^2 \, \p_-\theta_2 \p_+\theta_2 \hspace{-1pt}  + \hspace{-1pt} 2\lambda^2 \, \p_-\widetilde{\psi} \p_+\widetilde{\psi} \hspace{-1pt}  + \hspace{-1pt} 4\lambda^2\cos\theta_2 \, \p_-\widetilde{\psi} \p_+\phi_2  \nonumber \\
&\hspace{25pt} + \hspace{-1pt} 4\lambda^2 \bigl(\cos\theta_1 \, \p_- \phi_1 \hspace{-1pt} + \hspace{-1pt} \p_- \psi\bigr)\bigl(\cos\theta_2 \, \p_+ \phi_2 \hspace{-1pt} + \hspace{-1pt} \p_+ \widetilde{\psi}\bigr) \Bigr)\,. 
\end{align}\vspace{-15pt}}

\paragraph{Gauge fixing and background.} Recall that the model we are considering is invariant under the gauge transformation $g_r \mapsto g_r h$, $h \in U(1)$. In the parametrisation \eqref{Eq:Parametrisation} used above, this gauge symmetry simply becomes the translation $(\psi,\widetilde\psi) \mapsto (\psi+\eta,\widetilde\psi-\eta)$ with local parameter $\eta\in\R$. We now use this freedom to set $\widetilde{\psi} = 0$. Having fixed the gauge, we can then rewrite the action \eqref{Eq:ActionT11} as a $\s$-model on the coset $SU(2)\times SU(2) / U(1)$, with coordinate fields $y = (\theta_1,\theta_2,\phi_1,\phi_2,\psi)$. This defines the background metric $G_{ij}$ and background $B$-field $B_{ij}$, in terms of which the action reads
\begin{equation} \label{Eq:ActionSigma}
S = \frac{1}{2}\iint \text{d}x \, \text{d}t \ \bigl(G_{ij} + B_{ij}\bigr) \p_- y^i \p_+ y^j\,.
\end{equation}
Setting $\widetilde\psi=0$ in \eqref{Eq:ActionT11}, we read for the metric:
\begin{equation} \label{Eq:Metric}
\! ds^2 = G_{ij} \text{d}y^i \text{d}y^j = \lambda_1^2(\text{d}\theta_1^2 + \sin^2\theta_1 \, \text{d}\phi^2_1) + \lambda_2^2(\text{d}\theta_2^2 + \sin^2\theta_2 \, \text{d}\phi^2_2) + \lambda^2(\text{d}\psi + \cos\theta_1 \, \text{d}\phi_1 +  \cos\theta_2 \, \text{d}\phi_2)^2\,,
\end{equation}
while the $B$-field is given by
\begin{equation} \label{Eq:BField}
B = \frac{1}{2} B_{ij} \; \dd y^i \wedge \dd y^j = \lambda^2 (\dd\psi + \cos\theta_1\,\dd\phi_1) \wedge (\dd\psi + \cos\theta_2\,\dd\phi_2)\,.
\end{equation}
We recognise \eqref{Eq:Metric} as the metric of the so-called $T^{1,1}$ manifolds~\cite{Page:1984ae,Romans:1984an,Candelas:1989js}. More precisely, it defines a family of metrics, which depend on the three parameters $\lambda_1$, $\lambda_2$ and $\lambda$. Let us note that certain members of this family possess additional interesting geometrical properties. For instance, the choice $\lambda_1^2 = \lambda_2^2  = 3\lambda^2/2$ yields an Einstein metric, which is of importance in the gauge-string correspondence, see e.g.~\cite{Klebanov:1998hh}. As explained for a general group $G$ in the paragraph \ref{Par:GMM}, the case $\lambda_1=\lambda_2=\lambda$ yields the conformal model of~\cite{Guadagnini:1987ty}, which for the group $SU(2)$ considered here has been studied in~\cite{PandoZayas:2000he}, where it has been used to construct a pure NS-NS supergravity solution.\footnote{A parafermionic integrable deformation of this conformal $\s$-model on $T^{1,1}$ has been considered in~\cite{Georgiou:2019nbz}, by specifying to $SU(2)$ a class of models studied in~\cite{Georgiou:2016urf}. It would be interesting to investigate whether this model can be obtained from a construction similar to the one presented in this article.}

By construction, the model considered in this subsection is integrable for any metric in this family, \textit{i.e.} for all values of the parameters $\lambda_1$, $\lambda_2$ and $\lambda$. However, let us stress that this integrability also requires the presence of a $B$-field in the model, namely the $B$-field \eqref{Eq:BField} whose global prefactor $\lambda^2$ is then fixed by the choice of the metric (for other choices of this prefactor, the model is non-integrable, see subsections \ref{SubSec:EoM} and \ref{SubSec:Spinning}).

\paragraph{Lax connection.} As proven in subsection \ref{SubSec:LaxLimit}, the model under consideration possesses two independent Lax connections $\mathcal{L}_\pm$ and $\widetilde{\mathcal{L}}_\pm$, which characterise its integrability. Let us discuss their explicit expressions in terms of the coordinate fields $(\theta_1,\theta_2,\phi_1,\phi_2,\psi)$. As it turns out, instead of $\Lc_\pm(z)$, it will be simpler to describe its gauge transformation $\widehat{\Lc}_\pm(z) = h^{-1}\Lc_\pm(z)h + h^{-1}\p_\pm h$ with $h=\exp(-\psi I_3)$. Let us then write these Lax connections in terms of their components in the decompositions $\widehat{\mathcal{L}}_\pm = \widehat{\mathcal{L}}^a_{\pm} I_a$ and $\widetilde{\mathcal{L}}_\pm = \widetilde{\mathcal{L}}^a_{\pm} I_a$ along the basis $I_a = i\sigma_a/2$ of $\mathfrak{su}(2)$. From \eqref{Eq:LimitLax1}, using the parametrisation  \eqref{Eq:Parametrisation}, we get for $\widehat{\mathcal{L}}_\pm$:
\begin{equation*}
\widehat{\mathcal{L}}^{\,1}_{+} = \frac{\bigl(\lambda^2-\lambda_1^2\bigr)z}{\lambda^2z^2-\lambda_1^2} \sin\theta_1 \, \p_+\phi_1\,, \hspace{25pt} \widehat{\mathcal{L}}^{\,2}_{+} = \frac{\bigl(\lambda^2-\lambda_1^2\bigr)z}{\lambda^2z^2-\lambda_1^2} \, \p_+ \theta_1\,,
\end{equation*}
\begin{equation*}
\widehat{\mathcal{L}}^{\,3}_{+} = \frac{1}{\lambda^2z^2-\lambda_1^2} \bigl(\bigl(\lambda^2-\lambda_1^2\bigr) \cos\theta_1 \, \p_+\phi_1 - \lambda^2(z^2 -1) (\cos\theta_2 \, \p_+ \phi_2 + \p_+\psi)\bigr)\,,
\end{equation*}
together with
\begin{equation*}
\widehat{\mathcal{L}}^{\,1}_{-} = \frac{\sin\theta_1 \, \p_-\phi_1}{z}\,, \hspace{25pt} \widehat{\mathcal{L}}^{\,2}_{-} = \frac{ \p_- \theta_1}{z}\,, \hspace{25pt} \widehat{\mathcal{L}}^{\,3}_{-} = \cos\theta_1 \, \p_-\phi_1\,.
\end{equation*}
Similarly, for $\widetilde{\mathcal{L}}_\pm$ we get from \eqref{Eq:LimitLax2}:
\begin{equation*}
\widetilde{\mathcal{L}}^{\,1}_{+} = z\sin\theta_2 \, \p_+\phi_2\,, \hspace{25pt} \widetilde{\mathcal{L}}^{\,2}_{+} = -z \, \p_+ \theta_2\,, \hspace{25pt} \widetilde{\mathcal{L}}^{\,3}_{+} = -\cos\theta_2 \, \p_+\phi_2\,,
\end{equation*}
as well as
\begin{equation*}
\widetilde{\mathcal{L}}^{\,1}_{-} = -\frac{\bigl(\lambda^2-\lambda_2^2\bigr)z}{\lambda_2^2z^2-\lambda^2} \sin\theta_2 \, \p_-\phi_2\,, \hspace{25pt} \widetilde{\mathcal{L}}^{\,2}_{-} = \frac{\bigl(\lambda^2-\lambda_2^2\bigr)z}{\lambda_2^2z^2-\lambda^2} \, \p_- \theta_2\,,
\end{equation*}
\begin{equation*}
\widetilde{\mathcal{L}}^{\,3}_{-} = \frac{1}{\lambda_2^2z^2-\lambda^2} \bigl(\bigl(\lambda^2-\lambda_2^2\bigr)z^2 \cos\theta_2 \, \p_-\phi_2 + \lambda^2(z^2 -1) (\cos\theta_1 \, \p_- \phi_1 + \p_-\psi)\bigr)\,.
\end{equation*}

\subsection{Modification of the background, isometries and equations of motion}
\label{SubSec:EoM}

\paragraph{Isometries-preserving modification of the model.} Let us now consider a modification of the model described in the previous subsection (this will allow us to pinpoint the requirements for the integrability of the model and to make connections with other works in the next subsection). More precisely, let us take again an action of the form \eqref{Eq:ActionSigma}, with $y = (\theta_1,\theta_2,\phi_1,\phi_2,\psi)$ and metric given by \eqref{Eq:Metric}, but with the following $B$-field ($k \in \mathbb{R}$):
\begin{equation} \label{Eq:ModifBField}
B = k \, (\dd\psi + \cos\theta_1\,\dd\phi_1) \wedge (\dd\psi + \cos\theta_2\,\dd\phi_2)\,,
\end{equation}
obtained from \eqref{Eq:BField} by substituting the overall multiplication parameter $\lambda^2$ with $k$. For arbitrary values of $k$, this modification will break the integrability of the theory, while retaining the same isometries as the original model. In particular, as one can see from equations \eqref{Eq:Metric}, \eqref{Eq:BField} and \eqref{Eq:ModifBField}, the coordinates $\phi_1$, $\phi_2$ and $\psi$ do not appear in the coefficients of the metric and the $B$-field of both the original and the modified model and therefore the shifts $\phi_1 \to \phi_1 + \epsilon_1$, $\phi_2 \to \phi_2 + \epsilon_2$ and $\psi \to \psi + \epsilon$ are isometries of both backgrounds. 

For the original model, this is to be expected from the general results of section \ref{Sec:Model}. Indeed, as explained in subsection \ref{SubSec:GlobalSymm}, the model is invariant under the left translations $g_1 \mapsto f_1g_1$ and $g_2 \mapsto f_2g_2$, for $f_1,f_2 \in SU(2)$. In the parametrisation \eqref{Eq:Parametrisation}, the corresponding actions of the Cartan subgroup $\exp(\R I_3)$ of $SU(2)$ simply become shifts of the coordinates $\phi_1$ and $\phi_2$. Similarly, the shift of $\psi$ corresponds to the symmetry discussed in subsection \ref{SubSec:AdditionalSymm}. Consistently, $\phi_1$, $\phi_2$ and $\psi$ appear in the action \eqref{Eq:ActionT11} only through their derivatives.

One can calculate the Noether currents associated with these isometries for both models starting from the modified one. Following the conventions of appendix \ref{Sec:SpinningAnsatz}, we define the components of these currents as
\begin{equation}
\Pi^\mu_{\phi_1} = \frac{\p\Lc}{\p(\p_\mu \phi_1)}\,, \qquad \Pi^\mu_{\phi_2} = \frac{\p\Lc}{\p(\p_\mu \phi_2)} \qquad \text{ and } \qquad \Pi^\mu_{\psi} = \frac{\p\Lc}{\p(\p_\mu \psi)}\,,
\end{equation}
where $\mu$ are 2-dimensional space-time indices and $\Lc$ is the Lagrangian density of the action \eqref{Eq:ActionSigma}. In light-cone indices, one finds, using \eqref{Eq:Metric} and \eqref{Eq:ModifBField}:
\begin{subequations}\label{Eq:Currents}
\begin{align} 
2\Pi_{\phi_1}^\pm &= \left( \lambda_1^2 - (\lambda_1^2-\lambda^2) \cos^2\theta_1\right) \p_\mp \phi_1 + (\lambda^2 \mp k) \cos\theta_1 \bigl(\cos\theta_2 \, \p_\mp\phi_2 + \p_\mp\psi\bigr)\,, \\
2\Pi_{\phi_2}^\pm &= \left( \lambda_2^2 - (\lambda_2^2-\lambda^2) \cos^2\theta_2\right) \p_\mp \phi_2 + (\lambda^2 \pm k) \cos\theta_2 \bigl(\cos\theta_1 \, \p_\mp\phi_1 + \p_\mp\psi\bigr)\,, \\
2\Pi_{\psi}^\pm &= (\lambda^2 \pm k) \cos\theta_1 \, \p_\mp \phi_1 + (\lambda^2 \mp k) \cos\theta_2 \, \p_\mp \phi_2 + \lambda^2 \p_\mp \psi\,.
\end{align}
\end{subequations}
These Noether currents satisfy the conservation equations:
\begin{equation}\label{Eq:Conserv}
\p_\mu \Pi_i^\mu = \p_+ \Pi^+_i + \p_- \Pi^-_i = 0\,, \qquad \text{ for } \; i=\phi_1,\phi_2,\psi\,.
\end{equation}

\paragraph{Equations of motion.} Let us describe the equations of motion for the modified model. From the action \eqref{Eq:ActionSigma}, one obtains the following standard form:
\begin{equation*}
\p_- \p_+ y^i + \hat{\Gamma}^i_{jk} \, \p_- y^j \p_+ y^k = 0\,,
\end{equation*}
where $\hat{\Gamma}^i_{jk}$ are the components of the Christoffel symbol for the metric $G_{ij}$ modified by the torsion $T_{ijk}$ of the $B$-field $B_{ij}$, \emph{i.e.}
\begin{equation*}
\hat{\Gamma}^i_{jk} = \Gamma^i_{jk} - T^i_{jk} = \frac{1}{2}G^{im} \bigl(\p_j G_{mk} + \p_k G_{jm} - \p_m G_{jk} \bigr) - \frac{1}{2}G^{im} \bigl(\p_j B_{mk} + \p_k B_{jm} + \p_m B_{kj} \bigr)\,.
\end{equation*}
From \eqref{Eq:Metric} and \eqref{Eq:ModifBField}, we then find the following equations of motion for $\theta_1$ and $\theta_2$:
\small\begin{subequations} \label{Eq:EOM}
\begin{align}
\frac{\p_- \p_+ \theta_1}{\sin\theta_1} &= \p_- \phi_1 \left(\left(1 - \frac{\lambda^2}{\lambda_1^2}\right) \cos\theta_1 \, \p_+ \phi_1 - \frac{k+\lambda^2}{2\lambda_1^2} \bigl( \cos\theta_2 \, \p_+ \phi_2 + \p_+ \psi \bigr)\right) + \frac{k - \lambda^2}{2\lambda_1^2} \bigl( \cos\theta_2 \, \p_- \phi_2 + \p_- \psi \bigr)  \p_+\phi_1\,, \label{Eq:EOM1} \\
\frac{\p_- \p_+ \theta_2}{\sin\theta_2} &= \p_- \phi_2 \left(\left(1 - \frac{\lambda^2}{\lambda_2^2}\right) \cos\theta_2 \, \p_+ \phi_2 + \frac{k - \lambda^2}{2\lambda_2^2} \bigl( \cos\theta_1 \, \p_+ \phi_1 + \p_+ \psi \bigr)\right) - \frac{k+\lambda^2}{2\lambda_2^2} \bigl( \cos\theta_1 \, \p_- \phi_1 + \p_- \psi \bigr) \p_+ \phi_2\, . \label{Eq:EOM2}
\end{align}
\end{subequations}\normalsize
For simplicity and as we will not need them, we have omitted the equations for the isometric coordinates $\phi_1$, $\phi_2$ and $\psi$. However, one checks that they can be expressed as particular combinations of the conservation equations \eqref{Eq:Conserv} for the currents \eqref{Eq:Currents}.

\subsection{Spinning string solutions}
\label{SubSec:Spinning}

In this subsection, we describe a certain class of solutions of the equations of motion of the model with modified $B$-field \eqref{Eq:ModifBField}, obtained by a spinning string ansatz \cite{Frolov:2002av,Arutyunov:2003uj}. Note that spinning strings in $T^{1,1}$ manifolds (or closely related wrapped strings) have already been studied in~\cite{Kim:2003vn,Wang:2005baa,Basu:2011di,Basu:2011fw,Rigatos:2020hlq} in specific cases. In particular, the non-integrability of these solutions have been discussed in~\cite{Basu:2011di,Basu:2011fw,Rigatos:2020hlq}: we will compare our results with the ones of~\cite{Basu:2011di,Basu:2011fw,Rigatos:2020hlq} at the end of this subsection.

\paragraph{Spinning string ansatz.} We follow the procedure described in appendix \ref{Sec:SpinningAnsatz}, where we discuss the spinning string ansatz for a general $\s$-model with $B$-field. Since the model we are considering possesses three commuting isometries, in the coordinates $\phi_1$, $\phi_2$ and $\psi$, one can then search for spinning string solutions of the form:
\begin{equation} \label{Eq:SpinningAnsatz}
\qquad \theta_i = \theta_i(x)\,, \qquad \phi_i = \omega_i t + \widetilde{\phi}_i(x)\,, \qquad \psi = \psi(x)\,, 
\end{equation}
with $i\in\lbrace 1,2\rbrace$ and $\omega_1$ and $\omega_2$ constant parameters (more generally, one could also add a term $\omega t$ in the expression of $\psi$ as it is also an isometric coordinate: for simplicity, we will not consider this more general case here). The functions $\widetilde{\phi}_1(x)$, $\widetilde{\phi}_2(x)$ and $\psi(x)$ are the equivalent of the functions $\chi^j(x)$ in appendix \ref{Sec:Spinning}. As explained in this appendix, these functions are necessary to ensure the consistency of the ansatz. As we shall now see, they (or more precisely their derivatives) can be determined explicitly, which in the end will allow us to obtain ordinary differential equations governing the functions $\theta_1(x)$ and $\theta_2(x)$.

\paragraph{Equations of motion for the isometric coordinates.} As explained in appendix \ref{Sec:Spinning}, in the spinning string ansatz \eqref{Eq:SpinningAnsatz}, the spatial and temporal components of the Noether currents \eqref{Eq:Currents} do not depend on time, and therefore their conservation equations simply become
\begin{equation*}
\p_x \Pi_{\phi_1}^x =  \p_x \Pi_{\phi_2}^x =  \p_x \Pi_{\psi}^x = 0\,,
\end{equation*}
where $\Pi^x_i = \Pi^+_i - \Pi^-_i$. The above equations have solutions $\Pi_{\phi_1}^x =  \Pi_{\phi_2}^x = \Pi_{\psi}^x = 0$ if we choose the integration constant to be zero for simplicity. Using this, the derivatives of the functions $\widetilde{\phi}_1(x)$, $\widetilde{\phi}_2(x)$ and $\psi(x)$, which we denote with a dot as in appendix \ref{Sec:SpinningAnsatz}, can be solved for in terms of the functions $\theta_1(x)$ and $\theta_2(x)$. More precisely, applying the equation \eqref{Eq:dChi} in the present case, we get
\begin{subequations}\label{Eq:DerIso}
\begin{equation}
\dot{\widetilde{\phi}}_1 = -\frac{k}{\lambda_1^2} \, \omega_1\cot ^2 \theta_1\,, \qquad\qquad
\dot{\widetilde{\phi}}_2 = +\frac{k}{\lambda_2^2} \, \omega_2\cot ^2 \theta_2\,, \vspace{-8pt}
\end{equation}
\begin{equation}
\dot{\psi} = +\frac{k}{\lambda^2} \, \omega_1 \cos\theta_1 \left(1+\frac{\lambda^2}{\lambda_1^2} \cot^2\theta_1\right) - \frac{k}{\lambda^2} \, \omega_2 \cos\theta_2 \left(1+\frac{\lambda^2}{\lambda_2^2} \cot^2\theta_2\right).
\end{equation}
\end{subequations}

\paragraph{Equations of motion for the non-isometric coordinates and integrability.} Inserting the spinning string ansatz \eqref{Eq:SpinningAnsatz} and the expressions \eqref{Eq:DerIso} in the equations of motion \eqref{Eq:EOM1} and \eqref{Eq:EOM2} for the non-isometric coordinates, we get the following:
\begin{subequations}\label{Eq:EOMSpinning}
\begin{align} 
\ddot{\theta}_1 &= \omega _1 \sin\theta _1 \!\left(\omega _1\!\left(\left(\frac{\lambda ^2}{\lambda _1^2}-1\right)\! + \! \frac{k^2}{\lambda^2\lambda_1^2}\!\left(\left(1-\frac{\lambda ^2}{\lambda _1^2}\right) \!+\! \frac{\lambda^2}{\lambda_1^2\sin^4\theta_1}\right)\right) \cos \theta _1-\omega _2 \, \frac{k ^2-\lambda ^4}{\lambda ^2 \lambda _1^2}\cos\theta _2\right), \\
\ddot{\theta}_2 &= \omega _2 \sin\theta _2 \!\left(\omega _2\!\left(\left(\frac{\lambda ^2}{\lambda _2^2}-1\right)\! +\! \frac{k^2}{\lambda^2\lambda_2^2}\!\left(\left(1-\frac{\lambda ^2}{\lambda _2^2}\right) \!+\! \frac{\lambda^2}{\lambda_2^2\sin^4\theta_2}\right)\right) \cos \theta _2-\omega _1 \, \frac{k ^2-\lambda ^4}{\lambda ^2 \lambda _2^2}\cos\theta _1\right).
\end{align}
\end{subequations}
As justified for a general $\s$-model in appendix \ref{Sec:Spinning}, these are ordinary differential equations which involve only the functions $\theta_1(x)$ and $\theta_2(x)$ corresponding to the non-isometric directions of the background. For generic values of the parameters, these equations are coupled and we expect them to be non-integrable. This is consistent with the analysis carried out in \cite{Basu:2011di,Basu:2011fw}, where the authors consider wrapped strings solutions in the case $k = 0$ (\emph{i.e.} no $B$-field) and rule out integrability by proving that their motion is chaotic~\cite{Basu:2011di} or by using the theory of non-analytic integrability~\cite{Basu:2011fw}\footnote{More precisely, the works~\cite{Basu:2011di,Basu:2011fw} deal with a string model on $T^{1,1} \times $AdS$_5$, described by a Polyakov action. In this case, the equations of motion of the fields are supplemented with the Virasoro constraints coming from the worldsheet diffeomorphism invariance. The wrapped strings solutions considered in~\cite{Basu:2011di,Basu:2011fw} contain non-trivial dynamical degrees of freedom only in the $T^{1,1}$ part of the target space and more precisely in the coordinates $\theta_1$ and $\theta_2$. The equations obeyed by these coordinates are then the same as the ones obtained here for the $\s$-model on $T^{1,1}$ alone, \textit{i.e.} equations \eqref{Eq:EOMSpinning} with $k = 0$. Similar spinning strings solutions have also been studied in~\cite{Kim:2003vn}. Moreover, the analysis of~\cite{Basu:2011di,Basu:2011fw} was extended in~\cite{Rigatos:2020hlq} to the more general class of $L^{a,b,c}$ manifolds, which includes $T^{1,1}$.}. Yet, the general results of appendix \ref{Sec:AppIntegrability} show that starting from an integrable $\s$-model, for which the equations of motion can be recast as a zero curvature equation, and applying the spinning string ansatz to the latter, one will find (under certain assumptions) a Lax equation for the mechanical system describing the dynamical variables of the spinning string ansatz. In our case, we thus expect the equations \eqref{Eq:EOMSpinning} to be integrable if the $\sigma$-model we start with is integrable. As explained in the previous subsections, this requires the addition of a $B$-field with the right coefficient, namely $k = \lambda^2$. This has the effect of cancelling the coupling terms in \eqref{Eq:EOMSpinning}, hence leaving us with equations of motion of two decoupled 1d systems, which are then trivially integrable.

\section{Conclusions}
In this work we have applied the general framework of affine Gaudin models to construct a new class of integrable coset $\sigma$-models. 
These are models on the product of $N$ copies of a Lie group $G$ modulo the action of a diagonal $G^{(0)}_{\rm diag}$ gauge symmetry. 
For $N=2$ we have obtained the corresponding Lagrangian and recast it in terms of the $\Rc$-matrix suggesting a generalisation for the case of arbitrary $N$ and $T$.
In the limiting case of a three-parameter family we observed a connection to some conformal field theories defined on homogeneous spaces. Finally, for $G=SU(2)$
we have obtained new integrable sigma models on $T^{1,1}$ manifolds and discuss their spinning string solutions.

There is a number of interesting questions which deserve further study. First of all, it would be desirable to  prove that generic $(N,T)$-models have 
the Lagrangian that fit our conjectural form (\ref{Eq:ActionRef}) given in terms of the classical $\Rc$-matrix. We checked the validity of this conjecture up to $(N=3,T=3)$,
and also for $N=1$ and $T$ arbitrary \cite{Young:2005jv}, but further evidence 
is welcome. Also, it would be nice to find an independent field-theoretic derivation of (\ref{Eq:ActionRef}) which bypass doing the Legendre transform. 

It would be also interesting to quantise the integrable models constructed here and study the corresponding renormalisation group flow. 
For instance, for the case of the integrable sigma model on $T^{1,1}$, it would be worth checking if the renormalisation flow preserves the form of the metric and the 
$B$-field allowing only the parameters $\lambda_1$, $\lambda_2$ and $\lambda$ to flow, in particular to reach the fixed point corresponding to  the conformal field theory
of Guadagnini, Martellini and Mintchev.

Since our approach is applicable for both compact and non-compact groups, one can try to construct in a similar fashion 
an integrable sigma model on  Lorentzian spaces $W_{4,2}=SL(2,{\mathbb R})\times SL(2,{\mathbb R})/U(1)$, that can be viewed as non-compact analogues of $T^{1,1}$. 
The combined sigma model on the 10-d homogeneous space $W_{4,2}\times T^{1,1}$ should then have a special conformal point in the parameter space 
which would correspond to a critical NS-NS superstring background \cite{PandoZayas:2000he}. Deviations from this point would be then regarded as integrable deformations of the corresponding conformal field theory. 

Finally, it would be very interesting to generalise the present approach to construct integrable coset sigma models based on supergroups. For $N=2$ one 
obvious candidate to take for $G$ is the supergroup $PSU(1,1|2)$, that has $SL(2,{\mathbb R})\times SU(2)$ as its bosonic subgroup. 
One might speculate that the corresponding integrable sigma model could have a special point in the parameter space 
corresponding to a critical string background, this time with both NS-NS and R-R fluxes.

\section*{Acknowledgements} 
We would like to thank Arkady Tseytlin for useful comments. S. L. would like to thank Fran\c{c}ois Delduc, Marc Magro, Volker Schomerus and Beno\^{i}t Vicedo for interesting discussions.
This work is funded by the Deutsche Forschungsgemeinschaft (DFG, German Research 
Foundation) under Germany's Excellence Strategy -- EXC 2121 ``Quantum Universe" -- 390833306.

\begin{appendices}

\section{Coefficients in the form \texorpdfstring{\eqref{Eq:HamiltonianCoeff}}{(3.1)} of the Hamiltonian} \label{Sec:Coefficients}

In this appendix, we give explicit expressions for the coefficients $a_{rs}^{(k)}$ and $b_{rs}^{(k)}$, where $r,s = 1,2$ and $k = 0,1$, appearing in equation \eqref{Eq:HamiltonianCoeff}. For the coefficients $b_{rs}^{(k)}$, we have:
\begin{subequations}
\begin{equation*}
b_{rs}^{(0)} = c_{\bar{r}\bar{s}}^{(0)}\frac{2 K \left(2 z_{\bar{r}}^4 + \zeta _+^2 \left(z_r^2-3 z_{\bar{r}}^2\right)+\zeta _-^2 \left(2 \zeta _+^2-z_1^2-z_2^2\right)\right)}{\left(z_{\bar{s}}^2-z_s^2\right)^3}\,,
\end{equation*}
\begin{equation*}
b_{rs}^{(1)} = c_{r\bar{s}}^{(1)}\frac{2 K \left(z_1^2 z_2^2\left(z_1^2+z_2^2\right)-\zeta _+^2\left(z_1^4+z_2^4\right) + \zeta _-^2 \left(\zeta _+^2 \left(z_1^2+z_2^2\right)-2 z_1^2 z_2^2\right)\right)}{z_1 z_2 \left(z_{\bar{s}}^2-z_{s}^2\right)^3}\,,
\end{equation*}
\end{subequations}
where we introduced the notation $\bar{r} = 3 - r$ for $r = 1,2$ and where the coefficients $c_{rs}^{(k)}$ are defined in \eqref{Eq:Cs}. For the coefficients $a_{rs}^{(k)}$, we have:
\begin{subequations}
\begin{equation*}
a_{rs}^{(0)} = b_{\bar{r}\bar{s}}^{(0)}\frac{K \left(2 z_s^4+\zeta _+^2 \left(z_{\bar{s}}^2-3 z_s^2\right)+\zeta _-^2 \left(2 \zeta _+^2-z_1^2-z_2^2\right)\right)}{2 \left(z_r^2-z_{\bar{r}}^2\right)^3}\,,
\end{equation*} \vspace{-15pt}
\begin{align*}
a_{rs}^{(1)} &= \frac{(-1)^{r+s}c_{\bar{r}\bar{s}}^{(1)}}{z_1^2 z_2^2 \left(z_1^2-z_2^2\right)^6} K^2\bigl(z_1^2 z_2^2 \left(2 \zeta _+^2-z_1^2-z_2^2\right) \left(2 \zeta _+^2 \left(z_1^4-z_2^2z_1^2+z_2^4\right)-z_1^2 z_2^2 \left(z_1^2+z_2^2\right)\right) \\
&-\zeta _-^2 \left(2 \zeta_+^2-z_1^2-z_2^2\right) \left(\zeta _+^2\left(z_1^2+z_2^2\right) \left(z_1^4+z_2^4\right)-4 z_1^4 z_2^4\right)+\zeta _-^4 \left(\zeta _+^2 \left(z_1^2+z_2^2\right)-2 z_1^2 z_2^2\right)^2\bigr)\,.
\end{align*}
\end{subequations}

\section{Reformulation of the action}
\label{App:Reformulation}

In this appendix, we give an expression of the coefficients $\rho_{rs}^{(k)}$ and $\kay_r$ defined in~\eqref{Eq:Coefficients}, with $\kay_1=\kay$ and $\kay_2=-\kay$, in terms of residues of well-chosen functions. This will allow us to reformulate the action \eqref{Eq:Action} in a compact way.\\

We start with the definition \eqref{Eq:PhiPM} of the functions $\vp_\pm(z)$, which we restate here for the reader's convenience:
\begin{equation*}
\vp_+(z) = \frac{z^2-\ze^2_+}{(z^2-z_1^2)(z^2-z_2^2)} \qquad \text{ and } \qquad \vp_-(z) = \frac{z(z^2-\ze^2_-)}{(z^2-z_1^2)(z^2-z_2^2)}\,.
\end{equation*}
We recall that in section \ref{Sec:Model}, we have made the choice $z_1=1$ and $z_2=x$ for the parameters $z_1$ and $z_2$. Note that in terms of the functions $\vp_\pm(z)$, the twist function \eqref{Eq:TwistZeroes} of the model takes the factorised form
\begin{equation}\label{Eq:FactTwist}
\vp(z) = 2K \vp_+(z)\vp_-(z)\,.
\end{equation}
Let us also define the functions
\begin{equation*}
\alpha_0(z,w) = \frac{z}{z^2-w^2} \qquad \text{ and } \qquad \alpha_1(z,w) = \frac{w}{z^2-w^2}\,.
\end{equation*}
Using the expression \eqref{Eq:Coefficients} of the coefficients $\rho_{rs}^{(k)}$ and $\kay_r$, one checks that they satisfy
\begin{equation}\label{Eq:RhoRes}
\rho_{rs}^{(k)} - \frac{\delta_{rs}}{2}\kay_r = -4K \res_{w=z_s} \res_{z=z_r} \alpha_k(z,w)\vp_+(z)\vp_-(w)\,.
\end{equation}
Note that the order in which we take the residues in the above equation is important. Indeed, for the opposite order, we have
\begin{equation*}
\rho_{rs}^{(k)} + \frac{\delta_{rs}}{2}\kay_r = -4K \res_{z=z_r} \res_{w=z_s} \alpha_k(z,w)\vp_+(z)\vp_-(w)\,.
\end{equation*}

Let us relate these expressions to the $\Rc$-matrix \eqref{Eq:DefRMatrix}. The latter can be re-expressed in terms of the projections $C\ti{12}^{(kk)}$ of the Casimirs on the gradations $\g^{(k)}$ (see paragraph \ref{Sec:Conventions}) as
\begin{equation*}
\Rc^0\ti{12}(z,w) = \sum_{k=0}^1 \alpha_k(w,z)C\ti{12}^{(kk)}\,.
\end{equation*}
This shows that for any elements $X,Y$ in the Lie algebra $\g$, we have
\begin{equation*}
\kappa\ti{12}\Bigl( \Rc^0\ti{12}(w,z), X\ti{1}Y\ti{2} \Bigr) = \sum_{k=0}^1 \alpha_k(z,w) \kappa\bigl( X^{(k)}, Y^{(k)} \bigr)\,.
\end{equation*}
Using this result, and reinserting the equation \eqref{Eq:RhoRes} in the action \eqref{Eq:Action}, we can rewrite the latter as
\begin{equation*}
S = \sum_{r=1}^2 S_{\text{W}\hspace{-1pt}\text{Z}\hspace{-1pt}\text{W}\hspace{-1pt},\,\kay_r}[g_r] - 4K \iint \dd x \, \dd t\, \sum_{r,s=1}^2 \ \res_{z=z_r} \res_{w=z_s} \kappa\ti{12}\Bigl( \Rc^0\ti{12}(w,z)\vp_+(z)\vp_-(w), j_{+,r}\null\ti{1} \, j_{-,s}\null\ti{2} \Bigr)\,,
\end{equation*}
which is the equation \eqref{Eq:ActionRef1} announced in the main text. Note also that, using the property \eqref{Eq:LaxInterp} of the Lax connection $\Lc_\pm(z)$, this expression can be further rewritten as
\begin{equation*}
S = \sum_{r=1}^2 S_{\text{W}\hspace{-1pt}\text{Z}\hspace{-1pt}\text{W}\hspace{-1pt},\,\kay_r}[g_r] -4K \iint \dd x \, \dd t\, \sum_{r,s=1}^2 \ \res_{z=z_r} \res_{w=z_s} \kappa\ti{12}\Bigl( \Rc^0\ti{12}(w,z)\vp_+(z)\vp_-(w), \Lc_{+}(z)\null\ti{1} \, \Lc_{-}(w)\null\ti{2} \Bigr)\,.
\end{equation*}

\section{Spinning string ansatz for a \texorpdfstring{$\bm\s$}{sigma}-model with \texorpdfstring{$\bm B$}{B}-field} \label{Sec:SpinningAnsatz}

\subsection{Generalities}

\paragraph{$\bm\s$-models with $\bm{B}$-field.} Let us consider a $\s$-model with coordinate fields $y^1(x,t),\cdots,y^N(x,t)$, metric $G_{ij}=G_{ji}$ and $B$-field $B_{ij}=-B_{ji}$, whose action is then
\begin{equation}\label{Eq:sAction}
S[y^1,\cdots,y^N] = \frac{1}{2} \iint \dd x\,\dd t \,(G_{ij}+B_{ij})\p_- y^i\,\p_+y^{\,j}\,.
\end{equation}
We denote by $\Lc=\frac{1}{2}(G_{ij}+B_{ij})\p_- y^i\,\p_+y^{\,j}$ the corresponding Lagrangian density. Let us define:
\begin{equation*}
\Pi_i^\mu = \frac{\p\Lc}{\p(\p_\mu y^i)}\,,
\end{equation*}
so that
\begin{equation*}
\Pi_i^\pm = \frac{1}{2}(G_{ij} \mp B_{ij} )\p_\mp y^{\,j}\,.
\end{equation*}
In space-time coordinates $(t,x)$, this becomes
\begin{subequations}\label{Eq:Pi}
\begin{eqnarray}
\Pi_i^t = \Pi_i^+ + \Pi_i^- = G_{ij} \, \p_t y^{\,j} + B_{ij} \, \p_x y^{\,j}\,, \\
\Pi_i^x = \Pi_i^+ - \Pi_i^- = -G_{ij} \, \p_x y^{\,j} - B_{ij} \,\p_t y^{\,j}\,.
\end{eqnarray}
\end{subequations}
The Euler-Lagrange equations of the action \eqref{Eq:sAction} can then be written as
\begin{equation}\label{Eq:EL}
\p_\mu \Pi_i^\mu = \frac{\p\Lc}{\p y^i}\,,
\end{equation}
for all $i\in\lbrace 1,\cdots,N\rbrace$.

\paragraph{Isometries.} Let us now suppose that the $\s$-model possesses an isometry along the coordinate $y^i$, \textit{i.e.} that the metric $G_{ij}$ and $B$-field $B_{ij}$ do not depend explicitly on $y^i$. In this case, the derivative of $\Lc$ with respect to $y^i$ vanishes and the equation of motion \eqref{Eq:EL} of $y^i$ becomes the conservation equation
\begin{equation}\label{Eq:Conse}
\p_\mu \Pi_i^\mu = \p_t \Pi^t_i + \p_x \Pi^x_i = 0\,.
\end{equation}
In particular, the quantities $\Pi_i^t$ and $\Pi_i^x$ are identified as the components of the Noether current associated with the global symmetry $y^i \mapsto y^i + \epsilon$ of the model and the Noether charge
\begin{equation*}
\int \dd x\;\Pi^t_i
\end{equation*}
is conserved under time evolution.

\subsection{Spinning string ansatz} \label{Sec:Spinning}

\paragraph{The ansatz.} Let us consider the above $\s$-model with coordinates $y^1,\cdots,y^N$ and $M$ an integer number smaller than $N$. We will suppose that the model possesses $N-M$ commuting isometries along its coordinates $y^{M+1},\cdots,y^N$. Our goal in this subsection will be to search for particular classical solutions of the equations of motion \eqref{Eq:EL} of this $\s$-model, by introducing the following ansatz for the fields $y^1,\cdots,y^N$:
\begin{equation}\label{Eq:Ansatz}
\begin{array}{ccl}
y^i = y^i(x),& & \text{ for } 1 \leq i \leq M\,, \\
y^i = \omega_i\,t + \chi^i(x), & & \text{ for } M+1 \leq i \leq N\,,
\end{array}
\end{equation}
where $\omega_i$, $i\in\lbrace M+1,\cdots,N\rbrace$, are constant numbers and $y^1(x),\cdots,y^M(x),\chi^{M+1}(x),\cdots,\chi^N(x)$ are functions of the worldsheet space coordinate $x$ only. As we shall see, the $t$-dependence of this ansatz will completely drop out of the equations of motion, yielding a coherent set of equations on the functions $y^i(x)$ and $\chi^i(x)$, in the coordinate $x$.

The usual spinning string ansatz, see {\it e.g.} \cite{Frolov:2002av,Arutyunov:2003uj}, corresponds to the case where the functions $\chi^{M+1}(x)$, $\cdots,\chi^{N}(x)$ vanish. As we will see, because of the presence of the $B$-field $B_{ij}$, these functions will be necessary to obtain a coherent ansatz. Moreover, we will also show that the equations of motion of these functions $\chi^i(x)$ can be explicitly solved in terms of the remaining functions $y^1(x),\cdots,y^M(x)$, yielding in the end a coherent set of coupled ordinary differential equations on the latter (under a certain assumption on the metric). Such a generalisation of the spinning string ansatz was considered in~\cite{Arutyunov:2003za}.

As a general remark, let us start by recalling that the equations of motion \eqref{Eq:EL} are expressed in terms of the quantities $\Pi_i^\mu$ defined in the previous subsection. Inserting the ansatz \eqref{Eq:Ansatz} in the expression \eqref{Eq:Pi} of $\Pi_i^t$ and $\Pi^x_i$, we get
\begin{subequations}\label{Eq:PiAnsatz}
\begin{align}
\Pi^t_i &= + \sum_{j=1}^M B_{ij} \, \dot{y}^{\,j}(x) + \sum_{j=M+1}^N \bigl( G_{ij} \,\omega_j + B_{ij}\,\dot\chi^j(x) \bigr)\,,\\
\Pi^x_i &= - \sum_{j=1}^M G_{ij} \, \dot{y}^{\,j}(x) - \sum_{j=M+1}^N \bigl( B_{ij} \,\omega_j + G_{ij}\,\dot\chi^j(x) \bigr) \,,\label{Eq:PiAnsatzSig}
\end{align}
\end{subequations}
where the dot denotes the derivative with respect to $x$.

Let us recall that the only dependences of the spinning string ansatz \eqref{Eq:Ansatz} on the worldsheet time coordinate $t$ are in the coordinates $y^{M+1},\cdots,y^N$, corresponding to isometries of the model. Because of these isometries, the metric $G_{ij}$ and $B$-field $B_{ij}$ do not depend explicitly on the coordinates $y^{M+1},\cdots,y^N$, and thus on the time $t$ under the ansatz \eqref{Eq:Ansatz}. In particular, this shows that the quantities $\Pi_i^t$ and $\Pi^x_i$ obtained in equation \eqref{Eq:PiAnsatz} do not depend on $t$.

\paragraph{Equations of motion for the isometric coordinates \texorpdfstring{$\bm{y^{M+1},\cdots,y^N}$.}{Y(M+1),...,YN}} Let us first focus on the coordinates $y^{M+1},\cdots,y^N$. Since they correspond to the isometries of the model, their equations of motion take the form of conservation equations (see equation \eqref{Eq:Conse}) $\p_t \Pi^t_i + \p_x \Pi^x_i = 0$, for all $i\in\lbrace M+1,\cdots,N \rbrace$. Then, as $\Pi^t_i$ does not depend on $t$ in the spinning string ansatz (see previous paragraph), these conservation equations simply become $\p_x \Pi_i^x=0$. These are trivially solved by 
\begin{equation*}
\Pi^x_i = C_{i}\,, \qquad \text{ for all } i\in\lbrace M+1,\cdots,N \rbrace\,,
\end{equation*}
where $C_{M+1},\cdots,C_N$ are integration constants. From the expression \eqref{Eq:PiAnsatzSig} of $\Pi_i^x$, the above equation can be rewritten as
\begin{equation*}
\sum_{j=M+1}^N G_{ij}\,\dot\chi^j(x) = - \sum_{j=M+1}^N B_{ij}\,\omega_j - \sum_{j=1}^M G_{ij}\,\dot{y}^{\,j}(x) - C_i\,,
\end{equation*}
for all $i\in\lbrace M+1,\cdots,N \rbrace$. To be able to proceed further, and in the rest of this appendix, we shall make the following assumption:
\begin{center}
\textbf{Assumption:} We suppose that the $(N-M)\times(N-M)$ matrix $(G_{ij})_{M+1 \leq i,j \leq N}$ is invertible.
\end{center}
We will then denote by $(H^{ij})_{M+1 \leq i,j \leq N}$ its inverse. Let us briefly comment on this. In other words, this assumption means that we suppose the restriction of the metric to the isometric directions to be invertible. Although the full metric $(G_{ij})_{1 \leq i,j \leq N}$ is of course an invertible matrix, it is possible for its submatrix $(G_{ij})_{M+1 \leq i,j \leq N}$ to be non-invertible. However, in the examples considered in this article, this assumption will be satisfied. Using the inverse matrix $H$, we then solve the above equation for $\dot\chi^i(x)$:
\begin{equation}\label{Eq:dChi}
\dot\chi^{i}(x) = -\sum_{j=M+1}^N H^{ij} \left( \sum_{k=M+1}^N B_{jk}\,\omega_k + \sum_{k=1}^M G_{jk}\,\dot{y}^k(x) + C_j \right), \qquad  \text{ for all } i\in\lbrace M+1,\cdots,N \rbrace\,.
\end{equation}
In particular, this gives the solution of the equations of motion of $y^{M+1},\cdots,y^N$ in terms of explicit integrals (indeed, the right hand-side of equation \eqref{Eq:dChi} and in particular the matrix $H^{ij}$ do not depend on the $\chi^j(x)$'s, as the corresponding coordinates $y^{\,j}$ are isometries of the model).\\

Let us briefly comment on the relation of the present results with the usual spinning string ansatz for a model without $B$-field. As explained in the previous paragraph, this usual ansatz corresponds to taking $\chi^i(x)=0$ for $i\in\lbrace M+1,\cdots,N \rbrace$. In this case, one has to make another assumption on the metric for the ansatz to be consistent, which is to suppose that its components $G_{ij}$ vanish for $i\in\lbrace M+1,\cdots,N \rbrace$ and $j\in\lbrace 1,\cdots,M \rbrace$, \textit{i.e.} that there are no metric terms mixing the isometric coordinates $y^{M+1},\cdots,y^N$ with the non-isometric coordinates $y^1,\cdots,y^M$. Under this assumption and supposing that there is no $B$-field (or at least no $B$-field mixing together the isometric coordinates $y^{M+1},\cdots,y^N$), the quantities $\Pi^x_i$, for $i\in\lbrace M+1,\cdots,N \rbrace$, vanish (see equation \eqref{Eq:PiAnsatzSig}). The equations of motion $\p_x\Pi^x_i=0$ are then trivially satisfied, ensuring the consistency of the usual spinning string ansatz. It is clear that the presence of a $B$-field in the isometric directions $y^{M+1},\cdots,y^N$ introduces non-vanishing terms in the expression \eqref{Eq:PiAnsatzSig} of $\Pi_i^x$: in this case, the consistency of the equations of motion $\p_x\Pi^x_i=0$ then requires choosing non-zero $\chi^j(x)$'s, which is why we introduced these functions in the more general ansatz \eqref{Eq:Ansatz}.

Let us finally note that in the notation of this paragraph, the usual spinning ansatz corresponds to taking the integration constants $C_i$ to be zero, as it gives $\Pi_i^x=0$. It is also possible to choose these constants to be non-zero and thus introduce new parameters in the final spinning string equations of motion. However, the consistency of the ansatz then requires to also introduce non-zero functions $\chi^j(x)$, even in the absence of a $B$-field.

\paragraph{Equations of motion for the non-isometric coordinates \texorpdfstring{$\bm{y^1,\cdots,y^M}$.}{Y1,...,YM}} Let us now study the equations of motion of the coordinates $y^1,\cdots,y^M$. For that, we will use the following standard form of the field equations of a $\s$-model:
\begin{equation}\label{Eq:EoM}
\p_- \p_+ y^i + \Gh^i_{\;jk} \, \p_-y^{\,j} \, \p_+y^k = 0\,,
\end{equation}
where $\Gh^i_{\;jk}$ are the Christoffel symbols of the metric $G_{ij}$ modified by the torsion of the $B$-field $B_{ij}$:
\begin{equation*}
\Gh^i_{\;jk} = \Gamma^i_{\;jk} - T^i_{\;jk} = \frac{1}{2} G^{im} \Bigl( \p_j G_{mk} + \p_k G_{jm} - \p_m G_{jk} \Bigr) - \frac{1}{2}G^{im} \Bigl( \p_j B_{mk} + \p_k B_{jm} + \p_m B_{kj} \Bigr)\,.
\end{equation*}
Considering $i\in\lbrace 1,\cdots,M \rbrace$ and inserting the ansatz \eqref{Eq:Ansatz} in the equation of motion \eqref{Eq:EoM}, we get:
\begin{align}\label{Eq:EoMA}
\ddot{y}^{\,i}(x) + \sum_{j=1}^M\,\sum_{k=1}^M \Gh^i_{\;jk} \, \dot{y}^{\,j}(x)\,\dot{y}^{\,k}(x) + \sum_{j=M+1}^N\,\sum_{k=M+1}^N \Gh^i_{\;jk} \, (\dot\chi^{\,j}(x)-\omega_j)\,(\dot\chi^{\,k}(x)+\omega_k) \hspace{30pt} \notag \\
+\sum_{j=1}^M \sum_{k=M+1}^N \Gh^i_{\;jk} \, \dot{y}^{\,j}(x)\,(\dot\chi^{\,k}(x)+\omega_k) + \sum_{j=M+1}^N \sum_{k=1}^M \Gh^i_{\;jk} \, (\dot\chi^{\,j}(x)-\omega_j)\,\dot{y}^{\,k}(x) = 0\,.
\end{align}
The quantities $\Gh^i_{\;jk}$ are defined in terms of the metric $G_{ij}$ and $B$-field $B_{ij}$. As the latter do not depend explicitly on the isometric coordinates $y^{M+1},\cdots,y^N$, so does $\Gh^i_{\;jk}$. In particular, under the ansatz \eqref{Eq:Ansatz}, the quantities $\Gh^i_{\;jk}$  do not depend on the time coordinate $t$. The equation \eqref{Eq:EoMA} is thus a differential equation only in the variable $x$. Moreover, let us note that the functions $\dot\chi^{\,j}(x)$ appearing in this equation are expressed explicitly in terms of $y^1(x),\cdots,y^M(x)$ and their derivatives through equation \eqref{Eq:dChi}. Finally, reinserting this expression in the above equation, one gets Ordinary Differential Equations (ODEs) of the form
\begin{equation}\label{Eq:ODE}
\ddot{y}^{\,i}(x) + F^i\bigl(y^{\,j}(x),\dot{y}^{\,j}(x)\bigr) = 0\,, \qquad \forall \, i\in\lbrace 1,\cdots,M \rbrace\,,
\end{equation}
for some explicit functions $F^i\bigl(y^{\,j},\dot{y}^{\,j}\bigr)$. We thus get a coherent one-dimensional dynamical system on $y^1(x),\cdots,y^M(x)$.

Let us make a brief comment on the method. We used equation \eqref{Eq:dChi} to eliminate the functions $\chi^j(x)$ of the system. Equation \eqref{Eq:dChi} only allows to express $\chi^j(x)$ as integrals over $x$, which are thus ``non-local'' quantities in terms of the functions $y^1(x),\cdots,y^M(x)$. However, it is important to notice that in the above analysis, the functions $\chi^j(x)$ appeared in the system only through their derivatives $\dot\chi^j(x)$ (because $y^{M+1},\cdots,y^N$ are isometric coordinates), which ensures that this replacement does not introduce any non-local terms in $y^1(x),\cdots,y^M(x)$. Thus, in the end, one really obtains an ODE of the form \eqref{Eq:ODE}, and not a non-local integro-differential equation.

\subsection{Integrability} \label{Sec:AppIntegrability}

If the $\s$-model we start from is integrable, a natural question is whether the induced 1d dynamical system \eqref{Eq:ODE} obtained from the spinning string ansatz is itself integrable. We investigate this question in this subsection. The integrability of the $\s$-model relies on the zero curvature equation
\begin{equation}\label{Eq:ZCE}
\p_x \Mc(z) - \p_t \Lc(z) + \bigl[ \Lc(z), \Mc(z) \bigr] = 0\,,
\end{equation}
of a Lax connection $\bigl(\Mc(z),\Lc(z)\bigr)$, depending on the spectral parameter $z\in\mathbb{C}$. In this subsection, we will make the following assumption on the Lax connection:
\begin{center}
\begin{minipage}{0.9\textwidth}
\textbf{Assumption:} The Lax connection $\bigl(\Mc(z),\Lc(z)\bigr)$ depends on the isometric coordinates $y^{M+1},\cdots,y^N$ only through their derivatives $\p_-^k \p_+^l y^i$ ($k+l>0$).
\end{minipage}
\end{center}
Let us comment briefly on this assumption. The zero curvature equation \eqref{Eq:ZCE} on $\bigl(\Mc(z),\Lc(z)\bigr)$ should be equivalent to the equations of motion of the $\s$-model \eqref{Eq:sAction}. The coordinates $y^{M+1},\cdots,y^N$ only enter these equations of motion through their derivatives $\p_-y^i$, $\p_+y^i$ and $\p_-\p_+y^i$, as they correspond to isometries of the model. Thus, the zero curvature equation \eqref{Eq:ZCE} involves only these derivatives. It is thus rather natural to expect that the Lax connection $\bigl(\Mc(z),\Lc(z)\bigr)$ itself also only depends on these derivatives. A subtlety in this reasoning is that the zero curvature equation \eqref{Eq:ZCE} is invariant under gauge transformations $\Mc(z) \mapsto h(z)^{-1}\Mc(z)h(z) + h(z)^{-1}\p_t h(z)$ and $\Lc(z) \mapsto h(z)^{-1}\Lc(z)h(z) + h(z)^{-1}\p_x h(z)$. In general, it is thus natural to expect that the Lax connection depends solely on the derivatives $\p_+y^i$, $\p_-y^i$ and $\p_+\p_-y^i$ only up to gauge transformations. If this is the case, one would then have to perform a gauge transformation to get to a Lax connection satisfying the above assumption.

We will now suppose that this assumption is verified and study the behaviour of the Lax connection under the spinning string ansatz \eqref{Eq:Ansatz}. For $i\in\lbrace M+1,\cdots,N\rbrace$, the derivatives $\p_-^k \p_+^l y^i$ take the form
\begin{equation*}
\p_-^k \p_+^l y^i = (\delta_{k0}\delta_{l1}+\delta_{k1}\delta_{l0})\omega_i + (-1)^k \frac{\dd^{k+l}\,}{\dd x^{k+l}}\chi^i(x)\,.
\end{equation*}
In particular, they do not depend on the worldsheet time coordinate $t$. As the non-isometric coordinates $y^1,\cdots,y^M$ do not depend on $t$ in the ansatz \eqref{Eq:Ansatz}, we thus conclude that the Lax connection $\bigl(\Mc(z),\Lc(z)\bigr)$ does not depend on $t$. In particular, the zero curvature equation \eqref{Eq:ZCE} then takes the form of the Lax equation of a mechanical system:
\begin{equation}\label{Eq:Lax}
\frac{\dd\;}{\dd x}\Mc(z) = \bigl[ \Mc(z), \Lc(z) \bigr]\,.
\end{equation}
This is not yet a Lax representation of the dynamical system \eqref{Eq:ODE}. Indeed, the matrices $\Mc(z)$ and $\Lc(z)$ still depend on the functions $\chi^i(x)$ and not only on the functions $y^i(x)$. However, because of the main assumption made in this subsection, they depend on these functions $\chi^i(x)$ only through their derivatives $\frac{\dd^k\,}{\dd x^k}\chi^i(x)$ ($k>0$, see above). These derivatives can be expressed in terms of the functions $y^i(x)$ through equation \eqref{Eq:dChi}. In then end, we then obtain an expression of the Lax pair $(\Mc(z),\Lc(z))$ in terms of the functions $y^1(x),\cdots,y^M(x)$ and their derivatives.

This is a good indication of the integrability of the spinning string system. Let us note however that in general, this does not ensure that the Lax representation \eqref{Eq:Lax} produces a sufficient number of conserved quantities, nor that these conserved quantities are in involution one with another (even if the field theory Lax connection one starts with satisfies a Maillet bracket). It seems difficult to address these questions in full generality. They would thus require a case by case analysis.

\end{appendices}

\end{document}